\documentclass[preprintnumbers,amsmath,amssymb,floatfix,11pt,prd,onecolumn,superscriptaddress,nofootinbib]{revtex4}
\usepackage{diagbox}
\usepackage{latexsym}
\usepackage{epsfig}
\usepackage{epstopdf,float}
\usepackage{graphicx}
\usepackage{amssymb}
\usepackage{amsmath}
\usepackage{dcolumn}
\usepackage{bm}
\usepackage{color}
\usepackage{comment}
\usepackage[shortlabels]{enumitem}
\usepackage{subfigure}
\usepackage{multirow}
\usepackage{xcolor} 
\begin{document}
\title{\bf Testing the Nature of Rotating Black Hole Shadows Surrounded by a Thin Accretion Disk within Rastall Gravity}

\author{Abdul Malik Sultan}
\altaffiliation{ams@uo.edu.pk, maliksultan23@gmail.com}
\affiliation{Department of Mathematics, University of Okara, Okara-56300 Pakistan}

\author{Muhammad Israr Aslam}
\altaffiliation{mrisraraslam@gmail.com, israr.aslam@umt.edu.pk}
\affiliation{Department of Mathematics, School of Science, University of Management and Technology, Lahore-54770, Pakistan.}

\author{Manahil Ali}
\altaffiliation{manomanahil600@gmail.com}
\affiliation{Department of Mathematics, University of Okara, Okara-56300 Pakistan}

\author{Dongping Su}
\altaffiliation{su\_dongping@outlook.com}\affiliation{School of Intelligent Engineering, Chongqing City Management College, Chongqing 401331, China}

\begin{abstract}
We investigate the observational appearance of a rotating black hole (BH) in Rastall gravity by analyzing its shadow and accretion signatures under different illumination environments. The spacetime geometry is characterized by the Rastall parameter $\mu$, the structure parameter $\gamma$, and the rotation parameter $a$. To visualize the BH environment, we employ a ray-tracing algorithm that follows photon trajectories from the observer's screen to the emission region. We analyze how the shadow radius, distortion, and photon ring morphology respond to changes in the spacetime parameters. For a fixed value of $a$, the shadow observables exhibit a pronounced dependence on the Rastall gravity parameters. In particular, increasing the structure parameter $\gamma$ leads to a gradual enlargement of the shadow radius, indicating an expansion of the photon capture region surrounding the BH. At the same time, the distortion parameter decreases, implying that the shadow boundary becomes progressively more circular and less deformed. These results suggest that larger values of $\gamma$ tend to suppress the asymmetry induced by rotation and enhance the apparent size of the shadow. Similar modifications are observed for different values of the Rastall parameter $\mu$, demonstrating that the combined effects of $\mu$ and $\gamma$ leave distinct signatures on the shadow morphology. Consequently, shadow observations may provide an effective tool for constraining the parameter space of rotating BHs in Rastall gravity.
 
\end{abstract}
\date{\today}
\maketitle
\section{Introduction}\label{intro}
The existence of BHs, once considered a purely theoretical prediction of gravitational physics, has now been strongly supported by observational evidence. The first detection of gravitational waves from merging BHs by the Laser Interferometer Gravitational-Wave Observatory (LIGO) provided strong, direct evidence for the physical existence of BHs. It established a new observational channel for studying compact objects under extreme gravitational conditions \cite{1}. A major observational achievement was recently achieved by the Event Horizon Telescope (EHT) observations by obtaining high-resolution images of the accretion structures around the supermassive BHs, such as M$87$ $^\ast$ and Sgr A$^\ast$, providing unprecedented details of their near-horizon regions \cite{2,3,4,5,6,7,8,9,10,11}. This remarkable observation suggests that the BH environment is illuminated by a magnetically arrested accretion disk, where strong magnetic fields regulate the inflow of matter \cite{12,13}. The dark central area in the image represents the BH shadow, produced by the trapping of photons within the strong gravitational field, whereas the surrounding bright ring, often called the photon ring, originates from light rays orbiting near the photon sphere before escaping toward the observer \cite{14}. These groundbreaking observations have established BH imaging as a powerful tool for probing space-time geometry around compact objects and testing modified theories of gravity.

Since its formulation by Einstein, General Relativity (GR) has been established as the fundamental framework for describing gravitational interactions, providing a geometrical interpretation of gravity through the curvature of space-time and successfully explaining a wide range of astrophysical and cosmological phenomena \cite{15}. Its predictions have been verified through various observational and experimental tests, including gravitational lensing, compact object dynamics, and the large-scale structure of the universe \cite{16}. However, despite its remarkable success, GR still faces several theoretical and observational challenges. At cosmological scales, it struggles to explain the late-time accelerated expansion of the universe without introducing exotic components such as dark energy \cite{17}, and it also fails to fully account for the unseen gravitational effects attributed to dark matter \cite{18}. Moreover, the theory predicts space-time singularities, such as those associated with BHs and the origin of the universe, where the classical description of gravity becomes invalid \cite{19}. These unresolved issues suggest that a more generalized framework of gravity may be required to achieve a deeper understanding of the gravitational interaction and cosmic evolution \cite{20}.

Over the last few decades, a broad class of modified gravitational theories has been developed to address several theoretical and observational challenges that remain unresolved within GR. In the study of BH physics, one of the fundamental concepts is the no-hair theorem, which states that, within the standard Einstein-Maxwell framework, an isolated BH can be fully described by only three conserved quantities, namely its mass, electric charge, and angular momentum, without supporting additional independent matter fields outside the event horizon \cite{21}. This idea, originally introduced by John Archibald Wheeler, played a central role in understanding the uniqueness and simplicity of BH solutions and has motivated extensive investigations into its validity beyond standard gravity. In this perspective, the no-hair conjecture has been widely examined in various alternative gravitational frameworks, including scalar–tensor models, Brans–Dicke theory, and the more general Horndeski theory \cite{22,23,24,25,26,27,28,29,30}. Several studies have shown that the classical no-hair theorem can be violated in extended gravitational frameworks, where BH solutions may support additional matter or scalar field configurations. Such examples include BHs coupled to dilaton, and Yang–Mills–Higgs fields, as well as scalarized BH solutions generated through non-minimal couplings between scalar fields and curvature invariants, including the Maxwell invariant, the Gauss–Bonnet gravity term, and the Ricci scalar \cite{31,32,33,34,35,36,36a}. In addition to these scalar-based extensions, several other modified gravity models, such as $f(R)$ gravity \cite{25}, $f(R,T)$ \cite{37} gravity, $f(Q)$ gravity \cite{38}, and Rastall gravity \cite{39}, have attracted considerable attention due to their rich gravitational structures and cosmological implications. These alternative frameworks provide broader possibilities for BH solutions and may lead to significant deviations from the predictions of standard GR, especially in the strong field regime.

A possible modification of GR can be achieved by reconsidering the standard conservation law of the energy–momentum tensor $\nabla_\alpha \mathcal{T}^{\alpha \beta}=0$, which is one of the fundamental assumptions of Einstein's gravity. However, in curved space-time, the validity of this conservation law may not remain strictly preserved under all physical conditions. Based on this idea, Rastall \cite{40} introduced an alternative gravitational theory in which the divergence of the energy–momentum tensor is no longer constrained to vanish identically, but instead acquires a direct dependence on the space-time geometry through a coupling parameter. This modification establishes a non-minimal interaction between matter and curvature, allowing the exchange of energy and momentum with the gravitational field. In the limiting case where the coupling parameter approaches zero, the theory smoothly reduces to the standard framework of GR. In particular, Rastall proposed that the non-conservation of the energy–momentum tensor is proportional to the gradient of the Ricci scalar, which leads to a modified gravitational field equation and provides a generalized description of gravitation in curved space-time.

\begin{eqnarray}\label{1}
R_{\alpha \beta}-\frac{1}{2}g_{\alpha \beta}R= \kappa(\mathcal{T}_{\alpha \beta}-\lambda g_{\alpha \beta} R).  
\end{eqnarray}
  Equation (\ref{1}) can be reformulated in the following equivalent form as
\begin{eqnarray}\label{2}
  R_{\alpha \beta}  +(\mu-\frac{1}{2}) g_{\alpha \beta}R=\kappa \mathcal{T}_{\alpha \beta},   
\end{eqnarray}
here, $\kappa$ represents the gravitational coupling constant of Rastall gravity, while the quantity $\mu= \kappa \lambda$ is commonly introduced as the Rastall coupling parameter, characterizing the strength of the matter–geometry interaction. An interesting feature of Rastall gravity is that all vacuum solutions obtained in GR remain valid within this theory, preserving consistency in the absence of matter sources. However, in the presence of matter fields, the resulting gravitational solutions explicitly depend on the Rastall coupling parameter and can differ considerably from their corresponding counterparts in GR \cite{41}. This dependence enriches the theoretical structure of Rastall gravity and provides a wider class of non-vacuum solutions with potentially distinct astrophysical and cosmological implications.

In recent years, Rastall gravity has attracted considerable attention as an alternative extension of GR, leading to a wide range of investigations in different gravitational and cosmological contexts. Its applications have been explored in standard cosmology \cite{42,43}, loop quantum cosmology \cite{44}, Kaluza–Klein theory \cite{45}, and Brans–Dicke theory \cite{46}. An important aspect of Rastall gravity is its possible phenomenological connection with particle creation mechanisms in curved space-time, where the exchange of energy between matter fields and space-time geometry naturally gives rise to the non-conservation of the stress–energy tensor \cite{47,48,49,50}. This feature provides an alternative interpretation for the modified conservation law and strengthens the physical motivation of the theory. In the context of BH physics, the first spherically symmetric BH solution surrounded by a perfect fluid in Rastall gravity was reported in \cite{41}. Following this development, several exact BH solutions have been constructed, including rotating BH configurations \cite{39,51}, Gaussian BHs \cite{52}, and non-commutative inspired BHs \cite{53}. A comparative analysis of BH solutions in Rastall gravity and their corresponding counterparts in GR can provide deeper insight into the effects of non-minimal matter–geometry coupling and its implications for strong gravitational phenomena. Within Rastall gravity, compact stellar models with physically acceptable stability and energy conditions have been constructed for astrophysical objects such as Vela $X-1$ and $4U1820-30$ \cite{53a}. Rastall gravity has also been applied in cosmological inflationary scenarios, including constant-roll warm inflation models, to investigate the effects of the Rastall coupling and dissipative parameters on inflationary observables and cosmic evolution \cite{53aa}. In addition, various wormhole solutions have also been investigated in Rastall-based gravitational frameworks, including models supported by anisotropic matter distributions, dark matter density profiles, and non-commutative geometries under conformal symmetries, where the geometrical behavior, energy conditions, stability criteria, and traversability properties of the resulting solutions were analyzed in detail \cite{53b,53c}. Moreover, the holographic Einstein ring of a charged Rastall AdS BH has been investigated in \cite{israr1}.

The analysis of BH shadows has become an effective theoretical and observational framework for exploring the properties of compact objects and testing gravitational theories in the strong-field regime. The shadow boundary is determined by unstable photon orbits around the BH and encodes essential information about the surrounding space-time geometry \cite{54}. In particular, shadow observables such as the radius, distortion, and angular diameter are highly sensitive to the physical parameters of the BH and the underlying gravitational model \cite{55}. Therefore, the study of photon dynamics and shadow formation provides a direct way to investigate the effects of modified gravity on the near-horizon structure of BHs. In the framework of Rastall gravity, the presence of non-minimal matter–geometry coupling can alter the space-time geometry and, consequently, modify the shadow features, offering a useful avenue for probing deviations from GR \cite{56}.

The study of the optical signatures of BH has become an important approach to understanding how light behaves in strong gravitational fields and to probing the underlying space-time geometry. In particular, the celestial light source framework provides an effective setup for analyzing BH shadows and the propagation of photons under uniform background illumination, allowing a clearer description of horizon-scale optical features \cite{57,58,59,60}. Earlier investigations have shown that several physical factors, including BH rotation, magnetic field effects, and the inclination angle of the observer, can significantly influence the morphology and observable characteristics of BH shadows, especially in rotating space-times \cite{61}. More recently, considerable attention has been devoted to studying the shadow properties of rotating BH \cite{62}, polarized imaging \cite{63}, quasinormal modes \cite{64}, nonlinear electrodynamic effects \cite{65}, thin and thick accretion disk models \cite{66,67}, shadow signatures in Lorentzian-Euclidean BHs \cite{battista1, last}, shadows of quantum Schwarzschild BH \cite{battista2} and thermodynamic aspects of BH solutions \cite{68,69} within different modified theories of gravity. These investigations indicate that alternative gravitational frameworks can provide a rich phenomenological platform for connecting theoretical corrections to observable astrophysical signatures. In the context of Rastall gravity, a comprehensive investigation of the optical properties of rotating BH remains relatively unexplored. Since space-time geometry directly determines the shadow structure and emission characteristics, such analyses can serve as a valuable observational probe for examining the effects of non-minimal matter geometry coupling beyond GR \cite{39}.

To explore these effects in detail, we employ a framework similar to that used in previous studies of BH imaging and accretion phenomena in modified gravitational theories \cite{70,71}. In particular, we consider a rotating BH solution in Rastall gravity and investigate the propagation of photons in the corresponding space-time by solving the null geodesic equations. The optical appearance of the BH is constructed through a backward ray-tracing method, where photon trajectories are traced from the observer’s image plane backward into the BH space-time, allowing an accurate determination of the shadow boundary and the associated emission profile. To model realistic astrophysical environments, we include physically relevant illumination scenarios, such as thin accretion disk emission and distant celestial light sources, and analyze how the observable image is influenced by important parameters, including the BH spin and the Rastall coupling parameter. Within this numerical setup, we evaluated key physical observables, such as the shadow characteristics and disk emission structure, and compared the obtained results with those predicted by GR and other modified gravity models. Since rotating BH configurations provide a more realistic description of astrophysical compact objects, and accretion disk models establish a direct connection with observational signatures, this analysis offers an effective way to examine the impact of Rastall coupling on BH observables and identify possible deviations from the standard predictions of GR. 

The paper is organized as follows. In Section \textbf{II}, we present the rotating BH solution in Rastall gravity and discuss the corresponding space-time geometry, photon motion, and shadow boundary. Section \textbf{III} is devoted to the analysis of BH shadow images under illumination from celestial light sources. In Section \textbf{IV}, we investigate the optical signatures of rotating BH in Rastall gravity and examine the impact of the Rastall coupling parameter and BH spin on the shadow structure, redshift distribution, and lensing features. We also employ observational data to constrain the model parameters and explore possible deviations from GR. Finally, Section \textbf{V} summarizes our main results and conclusions.

\section{Rotating Rastall Black Hole Solution}
In particular, Heydarzade and Darabi obtained a static and spherically symmetric BH solution surrounded by an anisotropic fluid distribution in Rastall gravity by \cite{41}. The corresponding spacetime line element is given by
\begin{eqnarray}\label{3}
 ds^2=-f(r)dt^2+f(r)^{-1}dr^2+r^2(d\theta^2+\sin^2\theta d\phi^2),  
\end{eqnarray}

here
\begin{eqnarray}\label{4}
  f(r)=1-\frac{2M}{r}-\frac{\gamma}{r^{\frac{1+3\omega_s-6\mu(1+\omega_s)}{1-3\mu(1+\omega_s)}}}.
\end{eqnarray}
The distribution of surrounding matter is described by an anisotropic energy-momentum tensor \cite{72}. In \cite{73}, the vacuum rotating BH solution in GR is described by the well-known Kerr metric, which is fully characterized by two fundamental parameters, namely the mass and angular momentum of the BH. The corresponding static BH solution in Rastall gravity given in Eq. (\ref{3}) \cite{41}, and it was later generalized to a rotating Kerr-like configuration in \cite{39,51}. By employing the modified Newman-Janis algorithm proposed by Azreg-A\"{i}nou \cite{74,75}, the rotating Rastall BH spacetime can be constructed from the static solution. The resulting geometry is described by four parameters: the BH mass $M$, rotation parameter $a$, surrounding field structure parameter $\gamma$, and the Rastall coupling parameter $\mu$ \cite{51}. In Boyer-Lindquist coordinates, the metric of the rotating Rastall BH takes the form \cite{51}
\begin{eqnarray}\nonumber
ds^2&=&-\bigg(1-\frac{2 M r +\gamma \, r^\nu}{\Sigma}\bigg)dt^2-\frac{2 a \sin^2\theta(2 M r +\gamma \, r^\nu)}{\Sigma}d\phi dt+\Sigma d\theta^2+ \frac{\Sigma}{\Delta}dr^2 \\ \label{5} &+& \sin^2 \theta \bigg(r^2+a^2+\frac{ a \sin^2\theta(2 M r +\gamma \, r^\nu)}{\Sigma}\bigg)d\phi^2,
\end{eqnarray}
along with
\begin{eqnarray}\nonumber
    \Delta&=& r^2+a^2-2Mr-\gamma\,r^\nu, \hspace{1cm} \Sigma=r^2+a^2\cos^2\theta, \\ \label{6} & & \nu= \frac{1-3\omega_s}{1-3\mu(1+3\omega_s)}.
\end{eqnarray}
In the limit $\gamma \rightarrow 0$, the metric given in Eq. \ref{5} reduces to the standard Kerr BH solution of GR \cite{73}, where $\omega_s$ represents the equation-of-state parameter of the surrounding fluid. For the particular case $\mu=0$ with $-1 < \omega_s \le -1/3$, the resulting geometry corresponds to a Kerr BH surrounded by quintessence matter \cite{76}. Moreover, the Schwarzschild spacetime can be recovered, when both $\gamma\rightarrow 0$ and $a\rightarrow 0$. For convenience, in the subsequent analysis we fixed $\omega_s=-1/3$. 
\begin{figure}
\centering
\subfigure[\tiny][~$\gamma=0.01$]{\label{a1}\includegraphics[width=5.4cm,height=5.2cm]{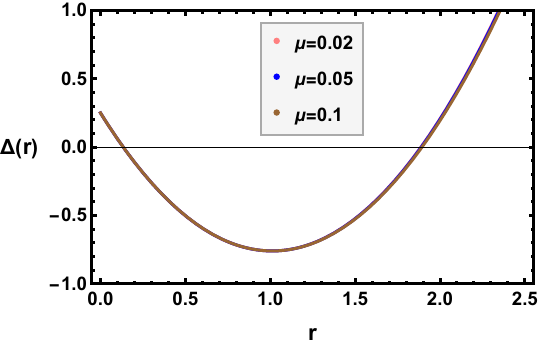}}
\subfigure[\tiny][~$\gamma=0.05$]{\label{b1}\includegraphics[width=5.4cm,height=5.2cm]{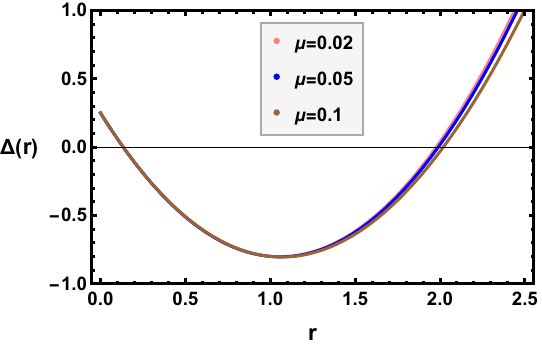}}
\subfigure[\tiny][~$\gamma=0.09$]{\label{c1}\includegraphics[width=5.4cm,height=5.2cm]{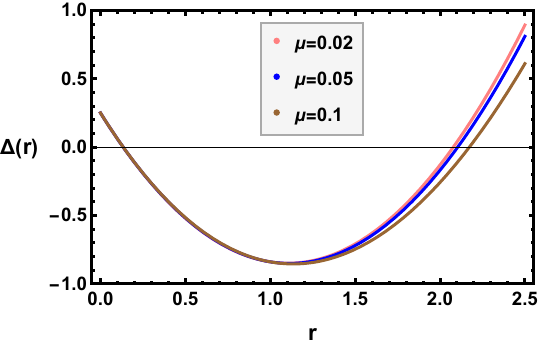}}
\subfigure[\tiny][~$\mu=0.02$]{\label{a1}\includegraphics[width=5.4cm,height=5.2cm]{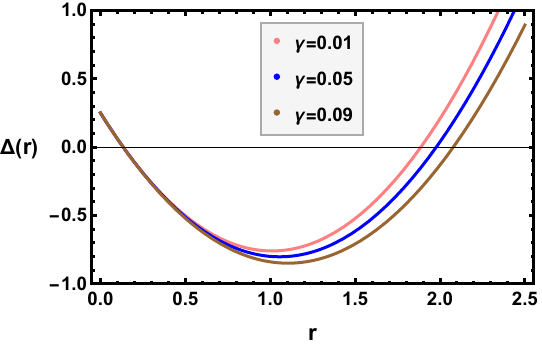}}
\subfigure[\tiny][~$\mu=0.05$]{\label{b1}\includegraphics[width=5.4cm,height=5.2cm]{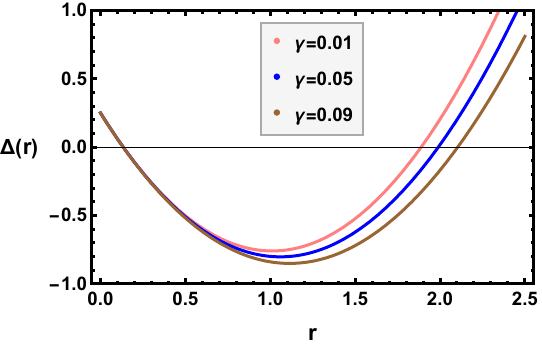}}
\subfigure[\tiny][~$\mu=0.1$]{\label{c1}\includegraphics[width=5.4cm,height=5.2cm]{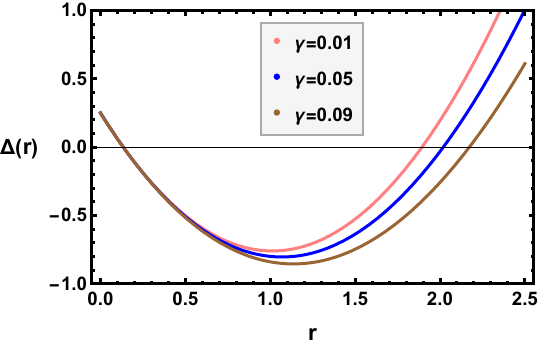}}
\caption{The physical behavior of the horizons for different values of the Rastall parameter $\mu$ and the structure parameter $\gamma$, with the BH spin parameter fixed at $a=0.5$. In the upper row, each panel corresponds to a fixed value of the structure parameter $\gamma$, while the pink, blue, and brown curves represent different values of the Rastall parameter $\mu$. In the lower row, each panel corresponds to a fixed value of the Rastall parameter $\mu$, whereas the pink, blue, and brown curves denote different values of the structure parameter $\gamma$.}\label{prd1}
\end{figure}

Figure \textbf{\ref{prd1}} depicts the behavior of the metric function $\Delta(r)$ as a function of the radial coordinate $r$ for different values of the Rastall parameter $\mu$ and the structure parameter $\gamma$, with the spin parameter fixed at $a=0.5$. The horizons are determined from the condition $\Delta(r)=0$, and the plots exhibit two distinct roots corresponding to the inner (Cauchy) horizon $r_{-}$ and the outer (event) horizon $r_{+}$, with $r_{-}<r_{+}$. In the upper row, each panel is plotted for a fixed value of $\gamma$ while varying $\mu$. It is observed that the separation between the curves increases slightly as the fixed value of $\gamma$ increases from one panel to another, indicating a stronger influence of the Rastall parameter on the horizon structure. In the lower row, each panel corresponds to a fixed value of $\mu$ with varying $\gamma$. In this case, the separation between the curves becomes progressively larger as the fixed value of $\mu$ increases, demonstrating that the impact of the structure parameter on the horizon radii is enhanced with higher values of the Rastall parameter. In general, both $\mu$ and $\gamma$ play a significant role in shaping the horizon structure and modifying the geometry of the BH spacetime.

The BH shadow arises from the propagation of photons in the strong gravitational field surrounding the BH. Photons travelling close to the photon sphere follow unstable null geodesic;, some are trapped by the BH, producing the dark shadow region, while others escape to a distant observer and delineate the shadow boundary, commonly referred to as the critical curve. Therefore, the characteristics of the observed shadow provide a powerful tool for probing the spacetime geometry and testing the effects of Rastall gravity in the strong-field regime. To investigate the motion of photons and massive particles in this background spacetime, we employ the Hamilton-Jacobi formalism, from which the geodesic equations can be derived as \cite{77}.
\begin{eqnarray}\label{7}
  \frac{\partial \mathcal{I}}{\partial \vartheta} = -\frac{1}{2} g^{\sigma \rho} \frac{\partial \mathcal{I}}{\partial x^{\sigma}} \frac{\partial \mathcal{I}}{\partial x^{\rho}},   
\end{eqnarray}
here, $\mathcal{I}$ denotes the Jacobi action associated with the photon trajectory, while $\vartheta$ is the affine parameter along the geodesic. Owing to the symmetries of the spacetime, the action $\mathcal{I}$ can be decomposed into the following separable form:
\begin{eqnarray}\label{8}
  \mathcal{I} = \frac{1}{2} \varsigma^2 \vartheta - E t + L \phi + S_r(r) + S_\theta(\theta),  
\end{eqnarray}
In the present analysis, we set $\varsigma=0$, corresponding to the case of massless particles (photons). The quantities $E=-p_t$ and $L=p_\phi$ are constants of motion associated with the spacetime symmetries, representing the photon's conserved energy and axial angular momentum, respectively. Furthermore, the functions $S_r(r)$ and $S_\theta(\theta)$ depend exclusively on the radial and angular coordinates. By inserting the separable form of the action into the Hamilton-Jacobi equation, one obtains the equations governing the photon geodesics.
\begin{eqnarray}\nonumber
  \Sigma^2 \frac{dt}{d\vartheta} &=& a \big(L - a E \sin^2\theta \big) + \frac{r^2 + a^2}{\Delta} \big[E(r^2 + a^2) - a L \big], \\ \nonumber 
 \Sigma^2 \frac{dr}{d\vartheta} &=& \pm \sqrt{\hat{R}(r)},\\ \nonumber
 \Sigma^2 \frac{d\theta}{d\vartheta} &=& \pm \sqrt{\Theta(\theta)}, \\ \label{9}
 \Sigma^2 \frac{d\phi}{d\vartheta} &=& \big(L \csc^2\theta - a E \big) + \frac{a}{\Delta} [E(r^2 + a^2) - a L],  
\end{eqnarray}
with
\begin{eqnarray}\nonumber
   \hat{R}(r) &=& [E(r^2 + a^2) - a L]^2 - \Delta [Q + (L - a E)^2 ], \\ \label{10}
   \Theta(\theta) &=& Q+ a^2 E^2 - L^2 \csc^2\theta \cos^2\theta.
\end{eqnarray}
Here, $Q$ represents the Carter constant, an additional conserved quantity arising from the separability of the geodesic equations. The above relations govern the propagation of photons in the BH spacetime. Circular photon trajectories define the photon sphere, located at a radius $r_{ps}$. For a photon to remain on such an orbit, the radial motion must satisfy the conditions $\dot{r}=0$ and $\ddot{r}=0$, where the overdot denotes differentiation with respect to the affine parameter $\vartheta$. These requirements are equivalent to imposing $\hat{R}(r_{ps})=0$ together with $\left. d\hat{R}/dr \right|_{r=r_{ps}}=0$. Under these conditions, the conserved quantities $E$, $L$, and $Q$ can be expressed in terms of the corresponding impact parameters that characterize photon trajectories in the vicinity of the BH.
\begin{eqnarray}\label{11}
\xi = \frac{L}{E}, \hspace{1cm} \zeta  = \frac{Q}{E^2}.  
\end{eqnarray}
Using Eq. (\ref{11}), one can derive the expressions for the associated impact parameters.
\begin{eqnarray}\label{12}
\xi(r_{ps})&=& \frac{(a^2+r_{ps}^2) \Delta' (r_{ps})-4 r_{ps} \Delta(r_{ps})}{a \Delta'(r_{ps})},   \\ \label{13} 
\zeta(r_{ps}) &=& \frac{r_{ps}^2 (-16\Delta(r_{ps})^2 - r_{ps}^2 \Delta'(r_{ps})^2 + 8\Delta(r_{ps})(2a^2 + r_{ps}\Delta'(r_{ps})))}{a^2 \Delta'(r_{ps})^2}.
\end{eqnarray}
In the above expressions, the prime symbol $'$ denotes differentiation with respect to the radial coordinate $r$. The photon region is obtained from the condition $\upsilon(r)=0$, whose roots correspond to the radii of the prograde and retrograde photon orbits and thereby determine the range of unstable circular photon trajectories. For physically admissible spherical photon motion, the angular function must additionally satisfy the constraint $\Theta(\theta)\geq 0$. To construct the BH shadow as perceived by a distant observer, we adopt the zero-angular-momentum observer (ZAMO) framework and map the photon trajectories onto the observer's image plane using a fisheye-camera representation together with stereographic projection techniques \cite{57,70}. The apparent shape of the shadow is then obtained through the relation between the photon's four-momentum and the celestial coordinates $(\epsilon,\upsilon)$ \cite{77}.
\begin{eqnarray}\label{14}
\cos\epsilon= \frac{p^{(1)}}{p^{0}} , \hspace{1cm}  \tan\upsilon= \frac{p^{(3)}}{p^{(2)}}.
\end{eqnarray}
To characterize the apparent image, we define Cartesian coordinates $(x,y)$ on the observer's plane, which are mapped to the associated celestial coordinates.
\begin{eqnarray}\label{15}
  x(r_{ps}) = -2 \tan\frac{\epsilon}{2}\sin\upsilon, \hspace{0.8cm} y(r_{ps}) = -2 \tan\frac{\epsilon}{2}\cos\upsilon.  
\end{eqnarray}

\begin{figure}
\centering
\subfigure[\tiny][~$\gamma=0.05$]{\label{a1}\includegraphics[width=7cm,height=7cm]{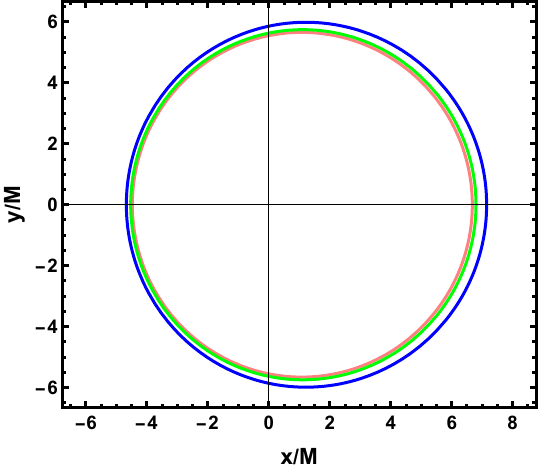}}
\subfigure[\tiny][~$\mu=0.05$]{\label{c1}\includegraphics[width=7cm,height=7cm]{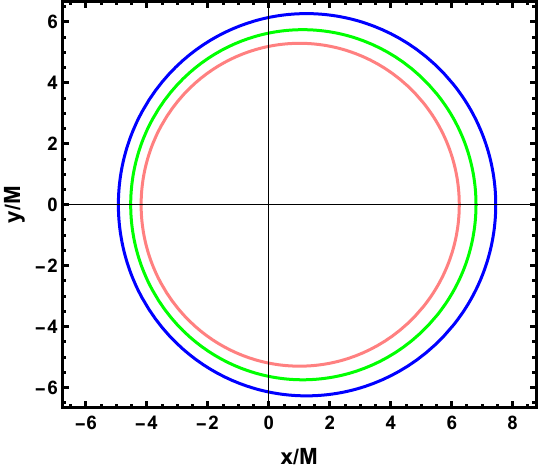}}
\caption{BH shadow profiles for a fixed spin parameter $a=0.5$. In the left panel, the structure parameter is fixed at $\gamma=0.5$, while the pink, green, and blue contours correspond to $\mu=0.02$, $0.05$, and $0.1$, respectively. In the right panel, the Rastall parameter is fixed at $\mu=0.05$, whereas the pink, green, and blue contours represent $\gamma=0.01$, $0.05$, and $0.09$, respectively. For both panels, the observer inclination angle is fixed at $\theta_{obs}=80^\circ$.}\label{prd2}
\end{figure}
To investigate the impact of the Rastall parameter $\mu$ and the structure parameter $\gamma$ on the shadow characteristics of the rotating BH, we consider several representative values of these parameters while keeping the spin parameter fixed at $a=0.5$. As illustrated in Fig.~\textbf{\ref{prd2}}, increasing either $\mu$ or $\gamma$ causes the shadow contour to shift toward the positive $x$-direction. At the same time, the radius of the shadow increases, leading to a larger shadow profile. The spacing between successive shadow contours also becomes more pronounced for larger parameter values, particularly in the case of increasing $\gamma$. Despite these variations in size and position, the shadow retains an almost circular shape, suggesting that the Rastall and structure parameters primarily influence the shadow radius and displacement, while having only a negligible effect on its overall deformation.

To quantitatively characterize the apparent shape of the rotating BH shadow, we employ two observational parameters: the shadow radius $R_d$, which measures the overall size of the shadow, and the distortion parameter $\delta_d$, which quantifies the deviation of the shadow boundary from a perfect circle. Using the definition presented in \cite{55}, the shadow observables are given by
\begin{eqnarray}\label{16}
  R_d= \frac{(x_t-x_r)^2+y_t^2}{2|x_t-x_r|} , \hspace{1cm} \delta_d= \frac{|x_{l'}-x_l|}{2 R_d}.
\end{eqnarray}
The overall size of the BH shadow is quantified through a reference circle of radius $R_d$, constructed using three characteristic points on the shadow boundary, namely the top, bottom, and rightmost points. To measure the departure of the shadow from an ideal circular shape, we introduce the distortion parameter $\delta_d$, defined by the horizontal displacement between the left edge of the shadow and the corresponding left edge of the reference circle. Here, $(x_t,y_t)$, $(x_b,y_b)$, $(x_r,0)$, and $(x_l,0)$ denote the uppermost, lowermost, rightmost, and leftmost points of the shadow contour, respectively, while $(x_l',0)$ represents the leftmost point of the reference circle. A perfectly circular shadow corresponds to the condition $x_l=x_l'$, yielding $\delta_d=0$. Any deviation from this equality results in a nonzero value of $\delta_d$, with larger values indicating a greater distortion of the shadow boundary from circularity.

\begin{figure}
\centering
\subfigure[\tiny][~$\mu=0.02$]{\label{a1}\includegraphics[width=7cm,height=7cm]{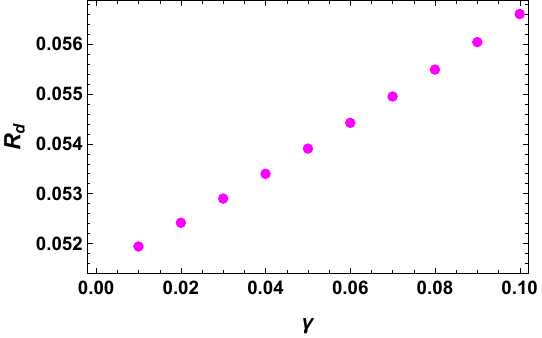}}
\subfigure[\tiny][~$\mu=0.05$]{\label{c1}\includegraphics[width=7cm,height=7cm]{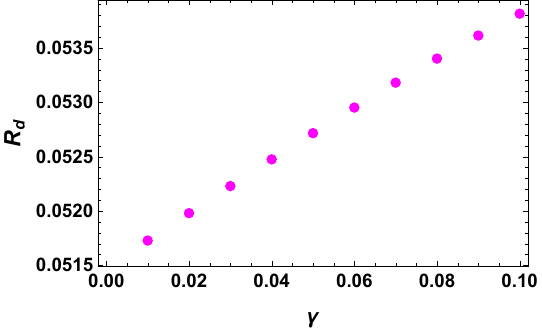}}
\subfigure[\tiny][~$\mu=0.02$]{\label{a2}\includegraphics[width=7cm,height=7cm]{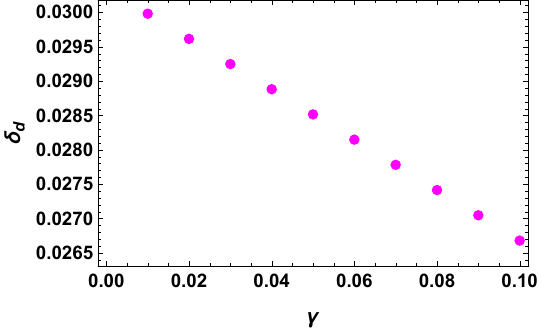}}
\subfigure[\tiny][~$\mu=0.05$]{\label{c2}\includegraphics[width=7cm,height=7cm]{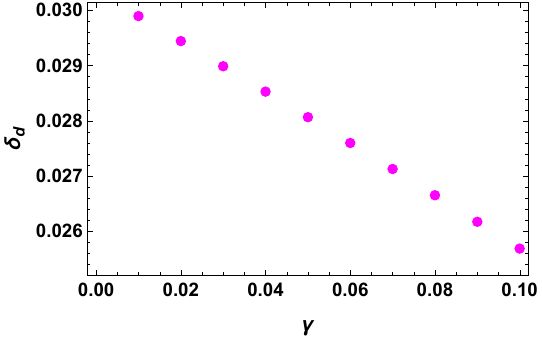}}
\caption{ The variation of the shadow observables $R_d$ and $\delta_d$ with the structure parameter $\gamma$ is presented for a fixed spin parameter $a=0.5$. The left and right panels correspond to $\mu=0.02$ and $\mu=0.05$, respectively. The observer is located at an inclination angle $\theta_{obs}=80^\circ$ and a radial distance $r_{obs}=100$.
}\label{prd3}
\end{figure}

The impact of the structure parameter $\gamma$ on the shadow observables $R_d$ and $\delta_d$ is examined in Fig. \textbf{\ref{prd3}} for a fixed spin parameter $a=0.5$. The left and right panels correspond to $\mu=0.02$ and $\mu=0.05$, respectively. It is observed that the shadow radius $R_d$ increases monotonically with increasing $\gamma$ in both cases, indicating an enlargement of the apparent shadow size. In contrast, the distortion parameter $\delta_d$ decreases as $\gamma$ increases, implying that the shadow gradually becomes less distorted. These results suggest that larger values of the structure parameter not only enhance the shadow radius but also drive the shadow toward a more circular and symmetric configuration. These findings are consistent with the shadow contours, as shown in Fig. \textbf{\ref{prd2}}.

\section{Shadow Formation under Celestial Illumination}
To investigate the visual appearance of the BH shadow in the presence of a celestial light source, we adopt a backward ray-tracing technique. In this approach, the BH shadow is represented as a dark circular region located at the center of the celestial sphere. The shadow radius is assumed to be much smaller than both the radius of the celestial sphere and the observer's distance from the BH, thereby maintaining a realistic observational setup. For a clearer visualization of photon trajectories, the celestial sphere is divided into four angular sectors, distinguished by the colors lavender, pink, orange, and yellow. This color coding enables the identification of the origin of photons reaching the observer. By tracing photon paths backwards from the observer to their emission points, the method selectively reconstructs only those trajectories that contribute to the observed image, providing an efficient framework for studying the optical signatures of BHs.

\begin{figure}
\centering
\subfigure[\tiny][~$\gamma=0.01,~\mu=0.02$]{\label{a1}\includegraphics[width=5.4cm,height=5.2cm]{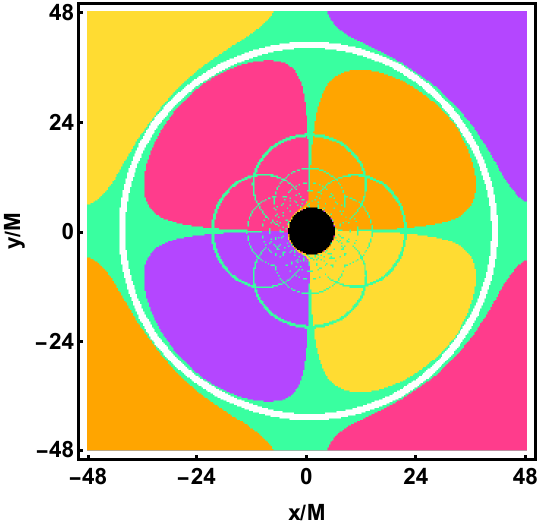}}
\subfigure[\tiny][~$\gamma=0.01,~\mu=0.05$]{\label{b1}\includegraphics[width=5.4cm,height=5.2cm]{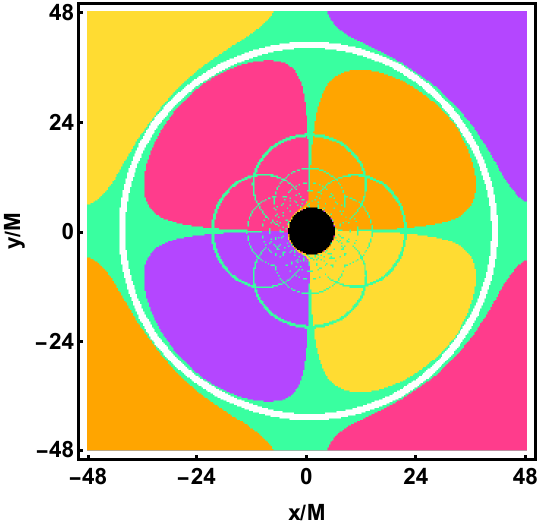}}
\subfigure[\tiny][~$\gamma=0.01,~\mu=0.1$]{\label{c1}\includegraphics[width=5.4cm,height=5.2cm]{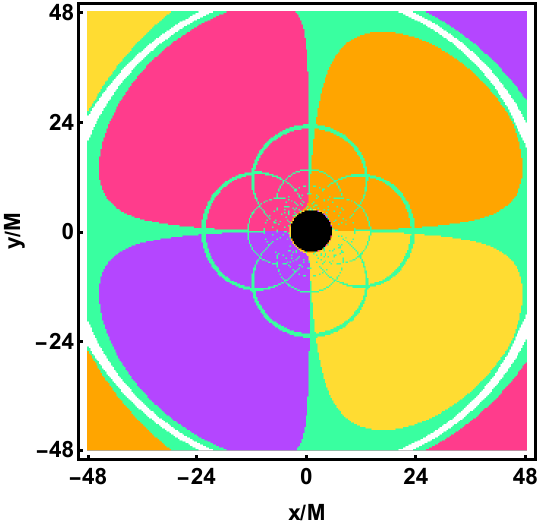}}
\subfigure[\tiny][~$\gamma=0.02,~\mu=0.02$]{\label{a1}\includegraphics[width=5.4cm,height=5.2cm]{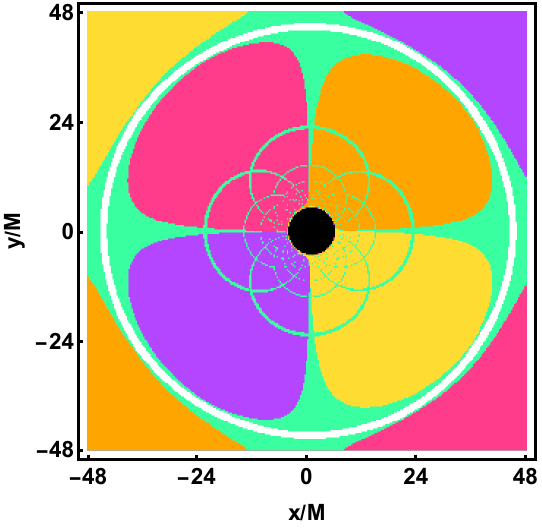}}
\subfigure[\tiny][~$\gamma=0.02,~\mu=0.05$]{\label{b1}\includegraphics[width=5.4cm,height=5.2cm]{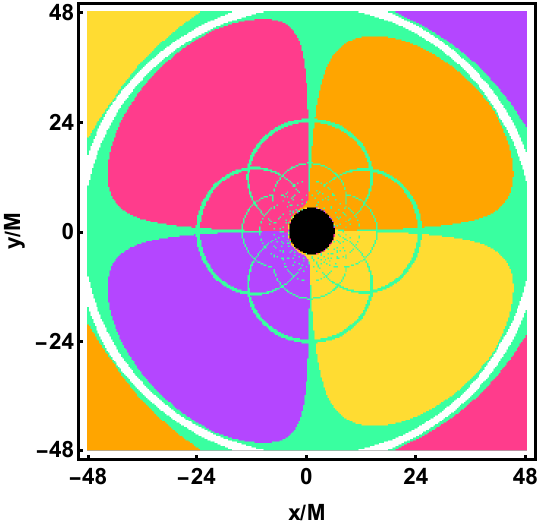}}
\subfigure[\tiny][~$\gamma=0.02,~\mu=0.1$]{\label{c1}\includegraphics[width=5.4cm,height=5.2cm]{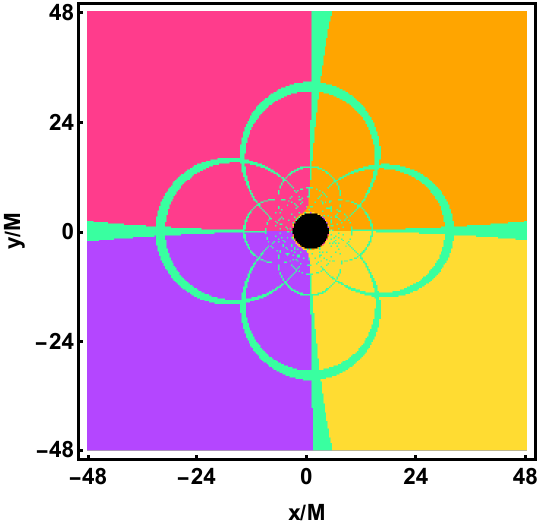}}
\subfigure[\tiny][~$\gamma=0.03,~\mu=0.02$]{\label{a1}\includegraphics[width=5.4cm,height=5.2cm]{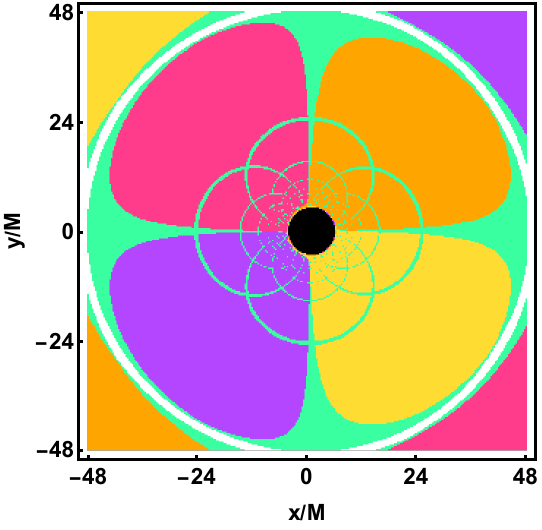}}
\subfigure[\tiny][~$\gamma=0.03,~\mu=0.05$]{\label{b1}\includegraphics[width=5.4cm,height=5.2cm]{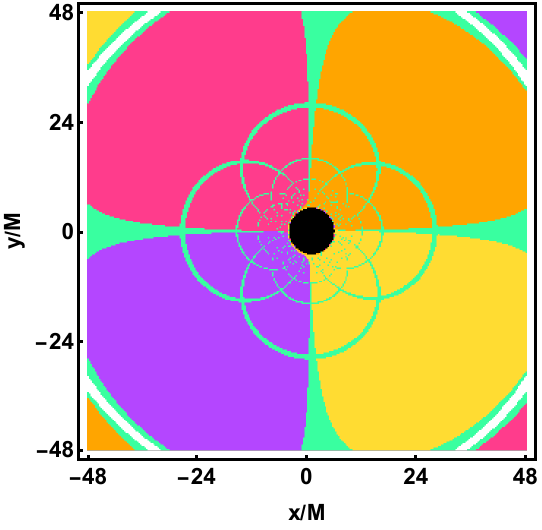}}
\subfigure[\tiny][~$\gamma=0.03,~\mu=0.1$]{\label{c1}\includegraphics[width=5.4cm,height=5.2cm]{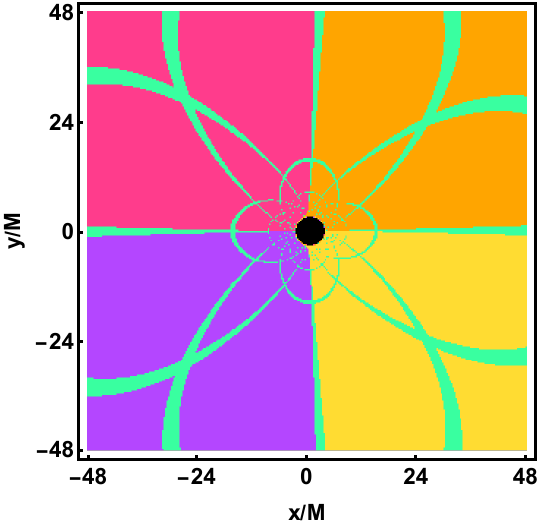}}
\caption{The BH shadow profiles produced by a celestial light source for different values of the Rastall parameter $\mu$ and the structure parameter $\gamma$ with fixed $a=0.5$ and $\theta_{ obs}=70^\circ$.
}\label{prd4}
\end{figure}

Following the procedure described in \cite{70}, we employ a fisheye camera projection model to account for observational effects in the imaging process. Using this framework, the BH shadow images are generated for different parameter values, as presented in the figure \textbf{\ref{prd4}}. In each panel, the central dark region corresponds to the BH shadow, while a bright circular ring surrounding it is identified as the Einstein ring, produced by the strong gravitational lensing of light. These image features provide a clear visualization of the influence of the BH gravitational field on photon trajectories, revealing the characteristic bending and focusing of light in the vicinity of the compact object. 

The shadow images corresponding to different values of  $\mu$ and $\gamma$ are shown in the figure \textbf{\ref{prd4}}, with the spin parameter fixed at $a=0.5$. In the first row, the structure parameter is fixed at $\gamma=0.01$, while the Rastall parameter varies as $\mu=0.02$, $0.05$, and $0.1$ from left to right. As $\mu$ increases, the Einstein ring expands outward, the central dark shadow region gradually decreases in size, and the characteristic D-shaped petals become more prominent. A similar behavior is observed in the second and third rows, corresponding to $\gamma=0.02$ and $\gamma=0.03$, respectively. Furthermore, for a fixed value of $\mu$, increasing $\gamma$ from top to bottom also enlarges the Einstein ring and the D-shaped petals while reducing the extent of the central shadow region. In the last column, corresponding to $\mu=0.1$, the Einstein ring becomes sufficiently large that portions of it extend beyond the image frame displayed for the higher values of $\gamma$. These results indicate that both $\mu$ and $\gamma$ have a significant impact on the apparent optical structure of the BH, primarily through enhancement of the Einstein ring and modification of the characteristics of the surrounding lens.
\section{Thin Accretion Disk Imaging}
Accretion disks around the BH provide a valuable framework for probing the physical properties of strong gravitational fields and the underlying spacetime geometry. In this section, we investigate the observational characteristics of thin accretion disks in the context of Rastall gravity and examine how the modified gravitational dynamics affect their observable signatures. The accreting material is modeled as a neutral plasma moving on equatorial timelike geodesics around the BH. The structure and emission properties of the disk are strongly influenced by several characteristic radii associated with the BH spacetime. Among these, the innermost stable circular orbit (ISCO) plays a central role, as it determines the inner boundary of the accretion disk and governs the efficiency of the accretion process. The location of the ISCO is closely related to the conversion of the rest-mass energy of infalling matter into radiation. Particles orbiting outside the ISCO follow stable circular trajectories and continuously contribute to the disk emission, whereas those crossing this radius become dynamically unstable and rapidly plunge toward the BH. Consequently, the radius $r_{ISCO}$ is determined by the conditions that define a marginally stable circular orbit \cite{71}.
\begin{eqnarray}\label{17}
    V_e(r)=0, \hspace{1cm} \partial_r V_e=0,\hspace{1cm} \partial_r^2V_e=0,
\end{eqnarray}
here, the quantity $V_e$ denotes the effective potential, which can be expressed as
\begin{eqnarray}\label{18}
V_e=(1+g^{tt}\hat{E}^{2}+g^{\phi \phi}\hat{L}^{2}-2g^{t\phi} \hat{E}\hat{L}).
\end{eqnarray}
The constants of motion associated with the particle trajectories are given by
\begin{eqnarray}\label{19}
\hat{E}=-\frac{1}{\sqrt{f_{1}}}(g_{tt}+g_{t\phi}\widetilde{\Psi}), \hspace{1cm}
\hat{L}=\frac{1}{\sqrt{f_{1}}}(g_{t\phi}+g_{\phi\phi}\widetilde{\Psi}),
\end{eqnarray}
here
\begin{eqnarray}\label{20}
  f_{1}=-g_{tt}-2g_{t\phi}\widetilde{\Psi}-g_{\phi\phi}\widetilde{\Psi}^{2},\hspace{0.5cm} \text{and} \hspace{0.5cm}
  \widetilde{\Psi}=\frac{d\phi}{dt}=\frac{\partial_{r}g_{t\phi}+\sqrt{\partial_{r}^{2}g_{t\phi}-\partial_{r}g_{tt}\,\partial_{r}g_{\phi\phi}}}{\partial_{r}g_{\phi\phi}}.
\end{eqnarray}
At $r=r_{ ISCO}$, the corresponding values of conserved energy and angular momentum are represented by $\hat{E}_{ ISCO}$ and $\hat{L}_{ISCO}$, respectively. For orbital radii satisfying $r>r_{ ISCO}$, the accreting material moves along stable Keplerian circular trajectories, and the associated four-velocity takes the form.
\begin{eqnarray}\label{21}
U^{\phi}_{out}=\frac{1}{\sqrt{f_{1}}}(1,0,0,\widetilde{\Psi}).
\end{eqnarray}
For radii smaller than the ISCO, the disk material departs from stable Keplerian motion and follows plunging trajectories directed toward the event horizon. Along these trajectories, the conserved quantities remain fixed at their ISCO values. The associated four-velocity components can therefore be written as \cite{71}.
\begin{eqnarray}\nonumber
U^{t}_{pl}&=&(-g^{tt}\hat{E}_{ISCO}+g^{t\phi}\hat{L}_{ISCO}),\quad
U^{\phi}_{pl}=(-g^{t\phi}\hat{E}_{ISCO}+g^{\phi\phi}\hat{L}_{ISCO}),\\\nonumber
U^{r}_{pl}&=&-\big(-(g_{tt}U^{t}_{pl}U^{t}_{pl}+2g_{t\phi}U^{t}_{pl}
U^{\phi}_{pl}+g_{\phi\phi}U^{\phi}_{pl}U^{\phi}_{pl}+1)(g_{rr})^{-1}\big)^{\frac{1}{2}},\\\label{22}~U^{\theta}_{pl}&=&0.
\end{eqnarray}
Photon trajectories may intersect the accretion disk one or more times before reaching the observer. A single intersection $(n=1)$ gives rise to the direct image, whereas two intersections $(n=2)$ produce the lensed image. Trajectories with $n>2$ correspond to higher-order images generated by photons undergoing multiple deflections around the BH. In this study, we focus exclusively on the direct and lensed components of the disk image. The observed radiation is primarily governed by the emission and absorption processes occurring within the accretion disk, while any contribution from reflected light is neglected. Under these assumptions, the specific intensity measured on the observer's image plane can be expressed as
\begin{eqnarray}\label{23}
K_{obs}=\sum_{n=1}^{N_{max}}f_{n}\psi_{n}^{3}(r_{n})\tau_{n}.    
\end{eqnarray}
Here, $K_{obs}$ denotes the photon frequency measured on the observer's image plane, while $N_{ max}$ specifies the maximum number of intersections between the photon trajectories and the accretion disk. The quantity $\psi_n$ represents the corresponding redshift factor, and $\tau_n$ characterizes the emissivity of the disk evaluated at the $n$th intersection point. The parameter $f_n=1$ is introduced as a normalization factor (fudge). The emissivity function $\psi_n$ is given by
\begin{equation}\label{24}
\tau_{n}=\exp\big[g_{1}k^{2}+g_{2}k\big],
\end{equation}
where $k=\log(\frac{r}{\hat{r}_h})$, while the parameters $g_1=-1/2$ and $g_2=-2$ are chosen in accordance with the $230$ GHz observations reported by the EHT \cite{78}. The quantity $\psi_n=\varepsilon_0/\varepsilon_n$ denotes the redshift factor, where $\varepsilon_0$ is the frequency measured by the distant observer and $\varepsilon_n$ is the frequency evaluated in the local comoving frame of the emitting matter. Because the emission properties of the accreting material differ inside and outside the ISCO, the corresponding redshift factors must be treated separately in these two regions. For photon emission originating outside the ISCO, the redshift factor takes the form \cite{79}
\begin{equation}\label{25}
\psi^{out}_{n}=\frac{\hat{e}_0(1-\Gamma\frac{p_{\phi}}{p_{t}})}{\rho_0(1+\widetilde{\Psi}\frac{p_{\phi}}{p_{t}})}|_{r=r_{n}},
\quad \quad r\geq r_{ISCO}.
\end{equation}
Here, $\hat{e}_0=\sqrt{\frac{g_{\phi\phi}}{g_{t\phi}^{2}-g_{tt}g_{\phi\phi}}},~ \Gamma=\frac{g_{t\phi}}{g_{\phi\phi}}$ and $\rho_0=\frac{1}{\sqrt{f_1}}$. The quantity $(\varrho)$, defined by
$[\varrho=\frac{p_{(t)}}{p_t}
=\hat{e}_0\left(1-\Gamma\frac{p_\phi}{p_t}\right),
]$ relates the energy measured by the observer to the conserved photon energy along a null geodesic. In an asymptotically flat spacetime, an observer located at spatial infinity measures a normalized energy $\varrho=1$. For radii smaller than the ISCO, however, the accreting matter no longer follows stable circular motion and instead plunges toward the BH along critical trajectories. The corresponding redshift factor in this plunging region can therefore be written as \cite{79}
\begin{equation}\label{23}
\psi^{pl}_{n}=-\frac{1}{U^{r}_{pl}p_{r}/p_{t}-\hat{E}_{ISCO}(g^{tt}-g^{t\phi}p_{\phi}/p_{t})
+\hat{L}_{ISCO}(g^{\phi\phi}p_{\phi}/p_{t}+g^{t\phi})}|_{r=r_{n}},
 \quad r< r_{ISCO}.
\end{equation}
\begin{figure}
\centering
\subfigure[\tiny][~$\gamma=0.01,~\mu=0.02$]{\label{a1}\includegraphics[width=5.4cm,height=5.2cm]{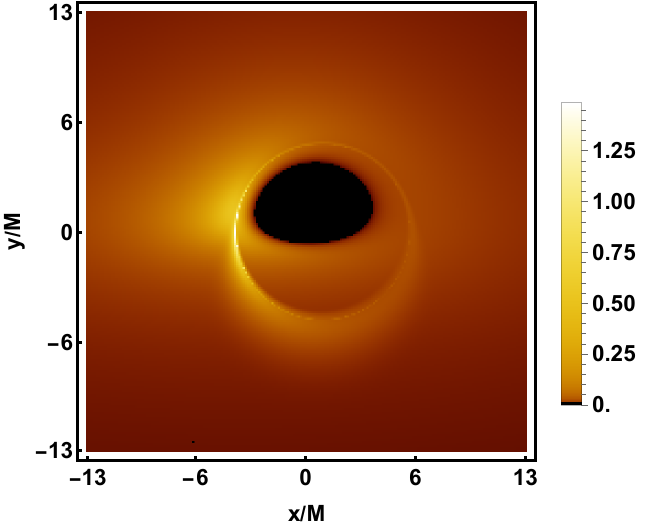}}
\subfigure[\tiny][~$\gamma=0.01,~\mu=0.05$]{\label{b1}\includegraphics[width=5.4cm,height=5.2cm]{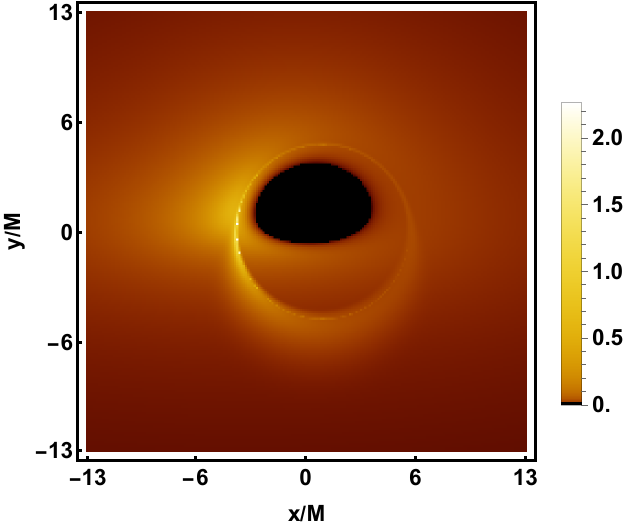}}
\subfigure[\tiny][~$\gamma=0.01,~\mu=0.1$]{\label{c1}\includegraphics[width=5.4cm,height=5.2cm]{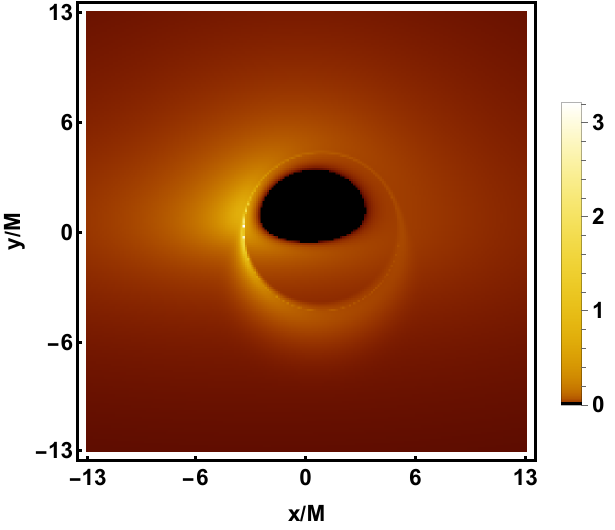}}
\subfigure[\tiny][~$\gamma=0.02,~\mu=0.02$]{\label{a1}\includegraphics[width=5.4cm,height=5.2cm]{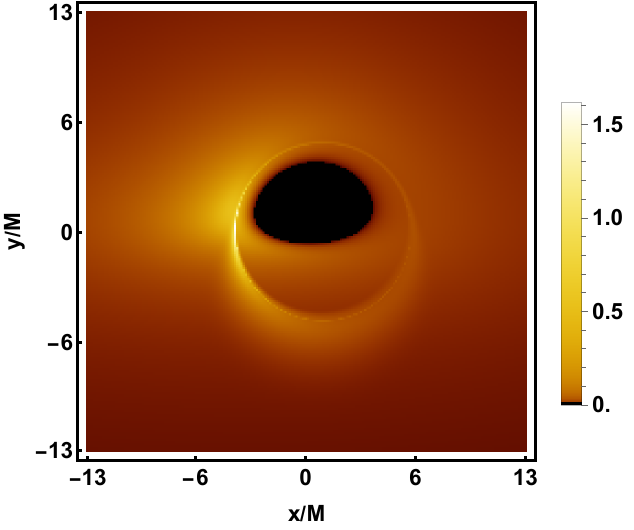}}
\subfigure[\tiny][~$\gamma=0.02,~\mu=0.05$]{\label{b1}\includegraphics[width=5.4cm,height=5.2cm]{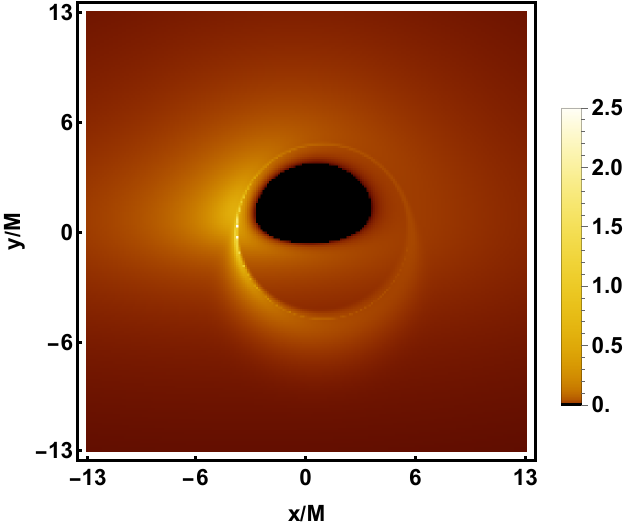}}
\subfigure[\tiny][~$\gamma=0.02,~\mu=0.1$]{\label{c1}\includegraphics[width=5.4cm,height=5.2cm]{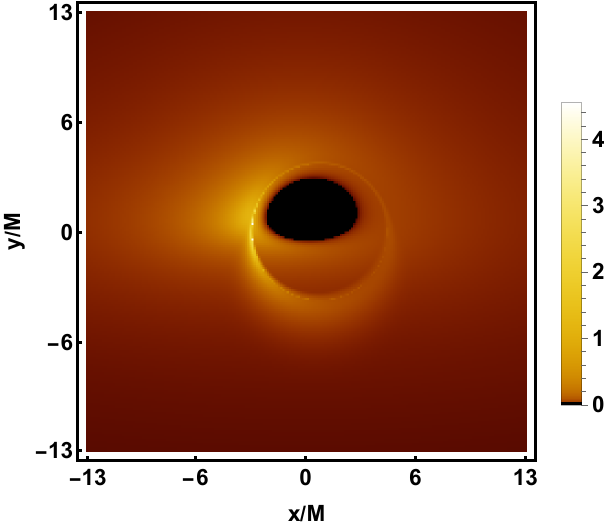}}
\subfigure[\tiny][~$\gamma=0.03,~\mu=0.02$]{\label{a1}\includegraphics[width=5.4cm,height=5.2cm]{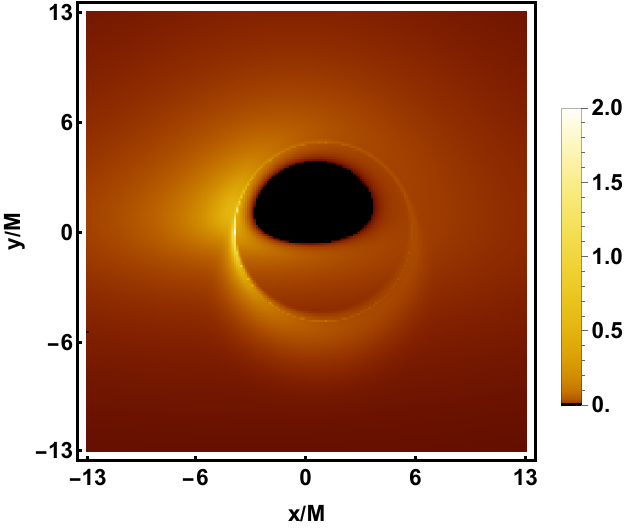}}
\subfigure[\tiny][~$\gamma=0.03,~\mu=0.05$]{\label{b1}\includegraphics[width=5.4cm,height=5.2cm]{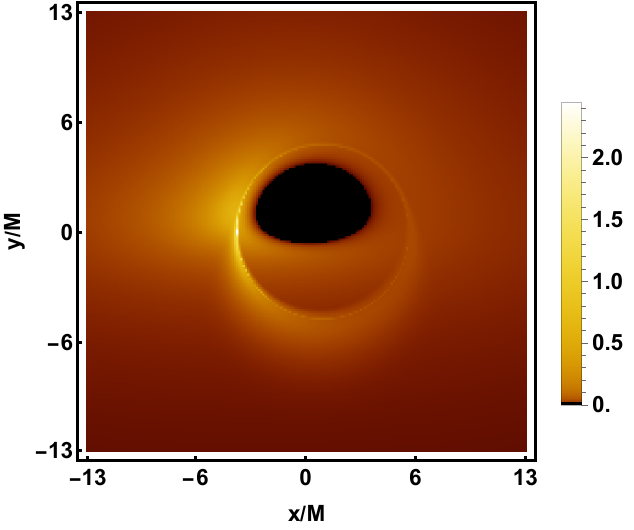}}
\subfigure[\tiny][~$\gamma=0.03,~\mu=0.1$]{\label{c1}\includegraphics[width=5.4cm,height=5.2cm]{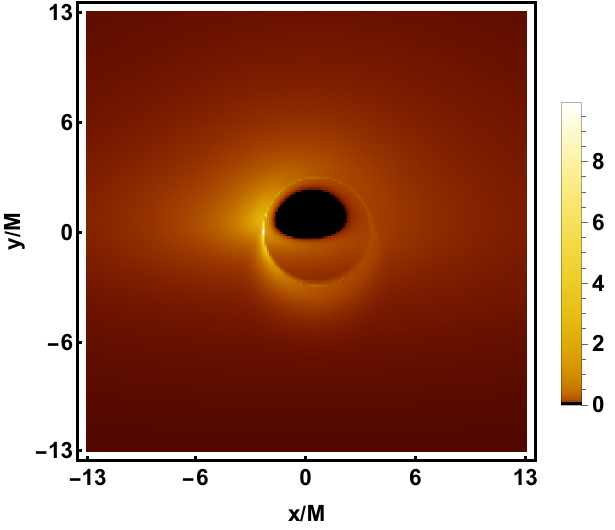}}
\caption{Optical images of a rotating BH in Rastall gravity for different values of $\mu$ and $\gamma$ with fixed $a=0.5$ and $\theta_{\rm obs}=70^\circ$ under a prograde accretion flow. The BH's event horizon is depicted as a black region and a luminous circular ring corresponds to the position of the photon ring.}\label{prd5}
\end{figure}
The optical appearance of the rotating BH in Rastall gravity under a prograde accretion flow is illustrated in Figure \textbf{\ref{prd5}} for different values of $\mu$ and $\gamma$, with the spin parameter fixed at $a=0.5$ and the observer inclination angle set to $\theta_{ obs}=70^\circ$. In all panels, a central dark region corresponding to the BH shadow is surrounded by an asymmetric emission profile produced by the accreting matter. The brightness is concentrated predominantly on one side of the image, a characteristic feature of prograde accretion flows arising from relativistic Doppler boosting. The Rastall parameter $\mu$ increases from left to right, whereas the structure parameter $\gamma$ increases from top to bottom. As $\mu$ increases, the shadow region gradually decreases, while the surrounding luminous structure expands outward. A similar trend is observed as $\gamma$ increases, indicating that both parameters significantly modify the apparent optical image of the BH. In addition, the lensing features around the shadow boundary become more extended for larger values of $\mu$ and $\gamma$, although the overall morphology of the image remains qualitatively unchanged. The largest apparent emission region and the smallest shadow size are obtained for the highest values of the $\mu$, demonstrating the influence of Rastall gravity on the observed intensity distribution around the rotating BH.

\begin{figure}
\centering
\subfigure[\tiny][~$\gamma=0.01,~\mu=0.02$]{\label{a1}\includegraphics[width=5.4cm,height=5.2cm]{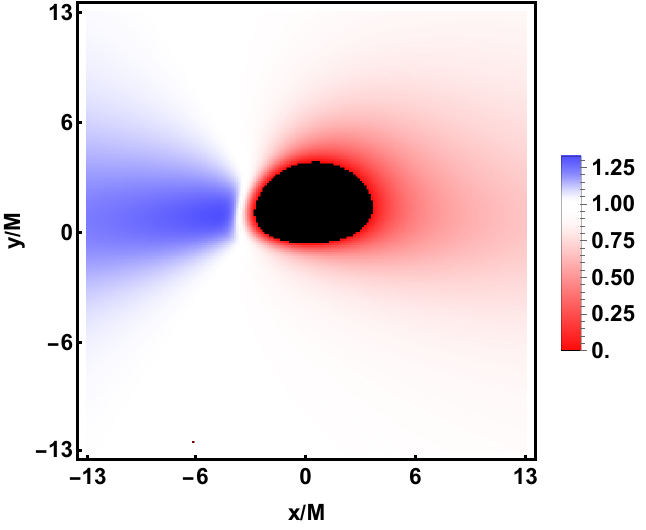}}
\subfigure[\tiny][~$\gamma=0.01,~\mu=0.05$]{\label{b1}\includegraphics[width=5.4cm,height=5.2cm]{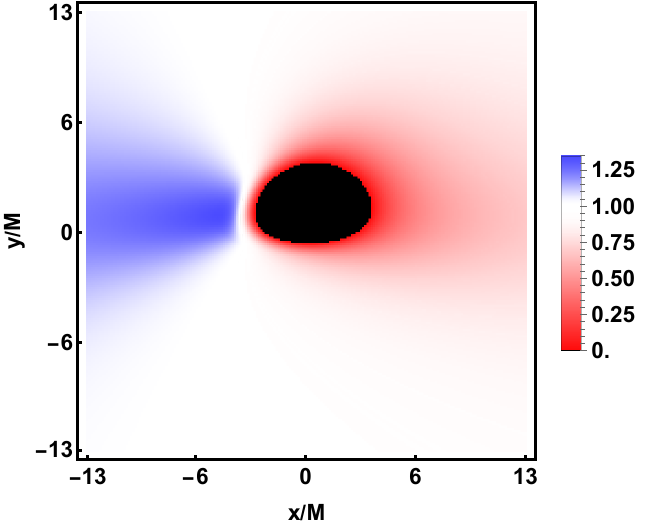}}
\subfigure[\tiny][~$\gamma=0.01,~\mu=0.1$]{\label{c1}\includegraphics[width=5.4cm,height=5.2cm]{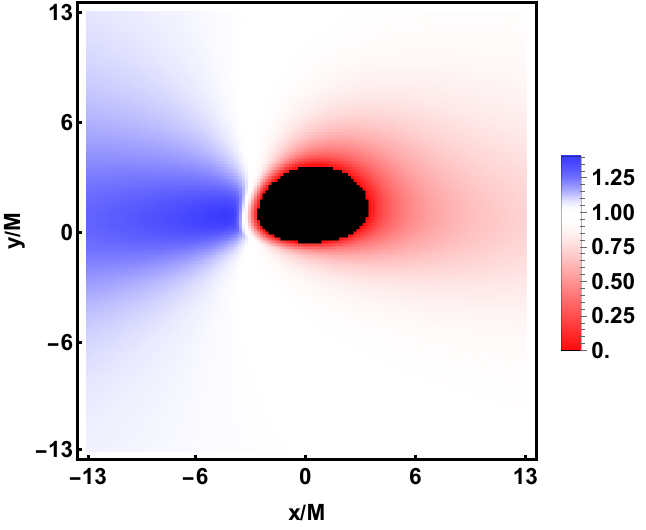}}
\subfigure[\tiny][~$\gamma=0.02,~\mu=0.02$]{\label{a1}\includegraphics[width=5.4cm,height=5.2cm]{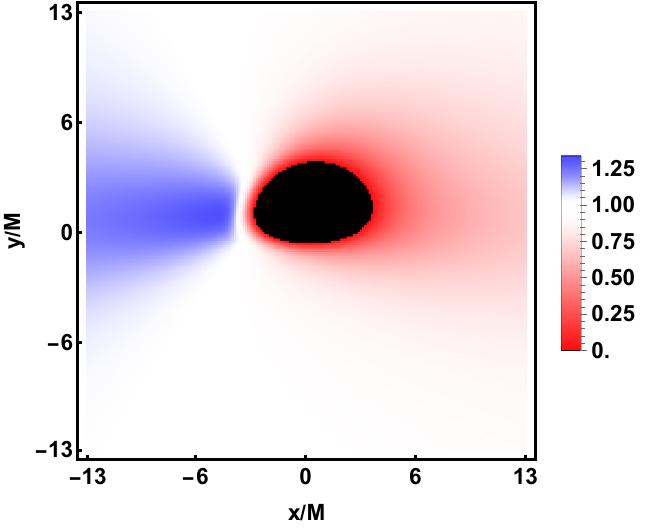}}
\subfigure[\tiny][~$\gamma=0.02,~\mu=0.05$]{\label{b1}\includegraphics[width=5.4cm,height=5.2cm]{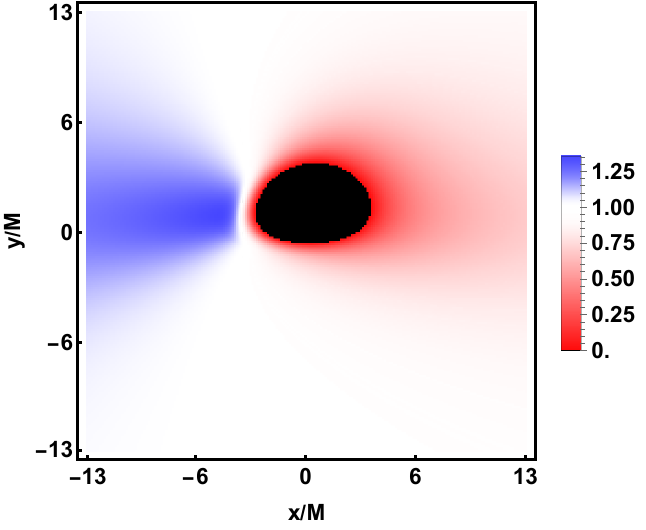}}
\subfigure[\tiny][~$\gamma=0.02,~\mu=0.1$]{\label{c1}\includegraphics[width=5.4cm,height=5.2cm]{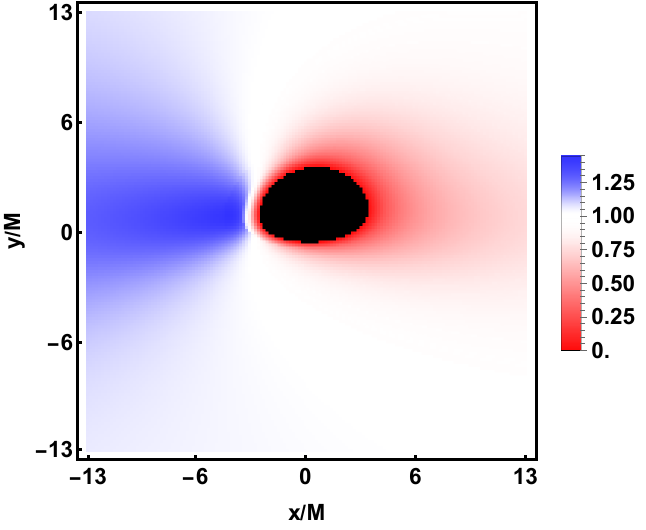}}
\subfigure[\tiny][~$\gamma=0.03,~\mu=0.02$]{\label{a1}\includegraphics[width=5.4cm,height=5.2cm]{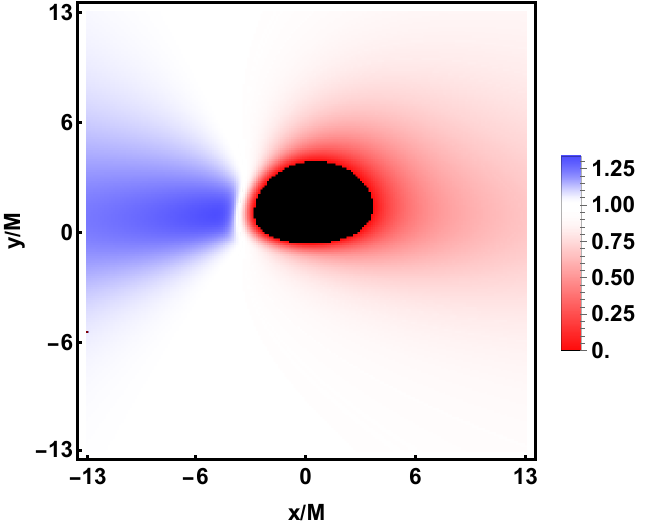}}
\subfigure[\tiny][~$\gamma=0.03,~\mu=0.05$]{\label{b1}\includegraphics[width=5.4cm,height=5.2cm]{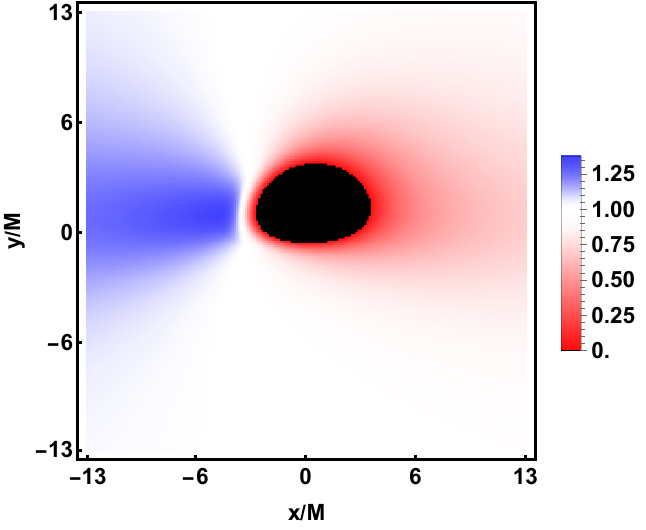}}
\subfigure[\tiny][~$\gamma=0.03,~\mu=0.1$]{\label{c1}\includegraphics[width=5.4cm,height=5.2cm]{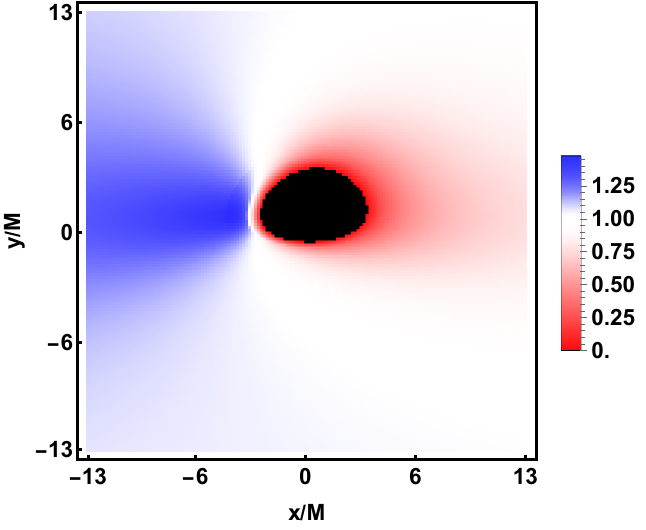}}
\caption{Redshift maps of the direct image for a rotating BH in Rastall gravity under a prograde accretion flow shown for different values of $\mu$ and $\gamma$ with fixed $a=0.5$ and $\theta_{ obs}=70^\circ$. The red and blue regions correspond to redshift and blueshift photon emission, respectively, while the black shaded area denotes the BH event horizon.
}\label{prd6}
\end{figure}
Figure \textbf{\ref{prd6}} presents the redshift distribution of direct images for a rotating BH in Rastall gravity with fixed spin parameter $a=0.5$. In the first row, the structure parameter is fixed at $\gamma=0.01$, while the Rastall parameter increases from left to right as $\mu=0.02$, $0.05$, and $0.1$. As $\mu$ increases, the blueshifted region gradually expands across the observer's screen, whereas the redshifted emission becomes comparatively more localized. In the second and third rows, corresponding to $\gamma=0.02$ and $\gamma=0.03$, respectively, the same qualitative behavior is observed. Moreover, for a fixed value of $\mu$, increasing the structure parameter from top to bottom further enhances the extent of the blueshifted region and alters the distribution of the redshifted emission. The most pronounced effect occurs in the last column, corresponding to $\mu=0.1$, where the blueshifted area occupies a substantial fraction of the image plane and becomes the dominant feature of the observed redshift pattern. The central dark region also exhibits noticeable variations with $\mu$ and $\gamma$. These results demonstrate that the Rastall and structure parameters significantly impact the redshift structure of the accretion flow and the overall appearance of the direct image.

\begin{figure}
\centering
\subfigure[\tiny][~$\gamma=0.01,~\mu=0.02$]{\label{a1}\includegraphics[width=5.4cm,height=5.2cm]{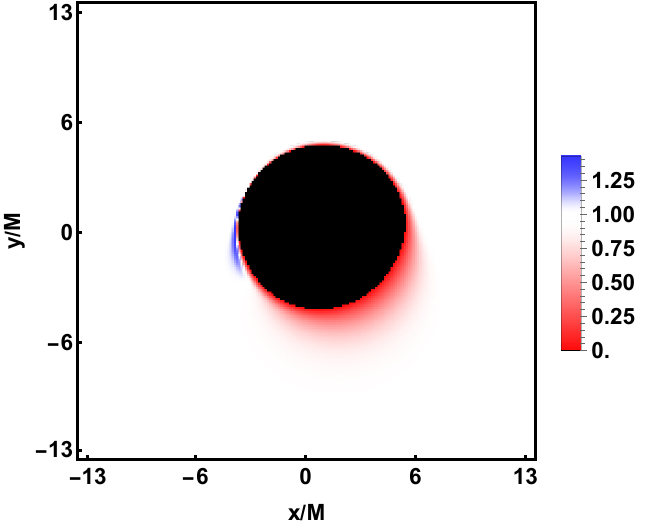}}
\subfigure[\tiny][~$\gamma=0.01,~\mu=0.05$]{\label{b1}\includegraphics[width=5.4cm,height=5.2cm]{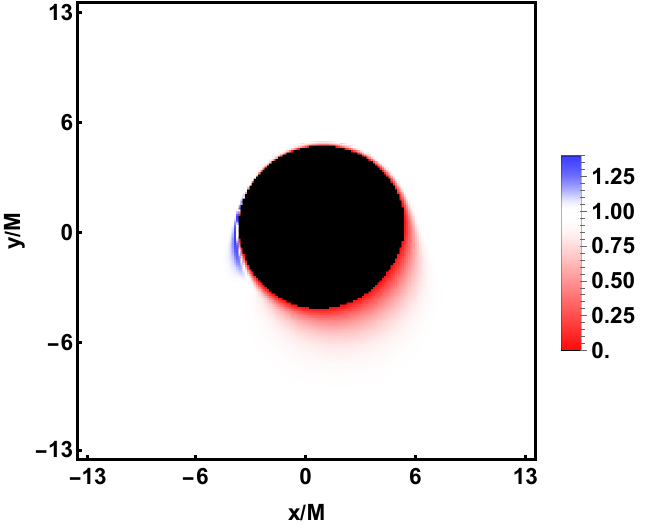}}
\subfigure[\tiny][~$\gamma=0.01,~\mu=0.1$]{\label{c1}\includegraphics[width=5.4cm,height=5.2cm]{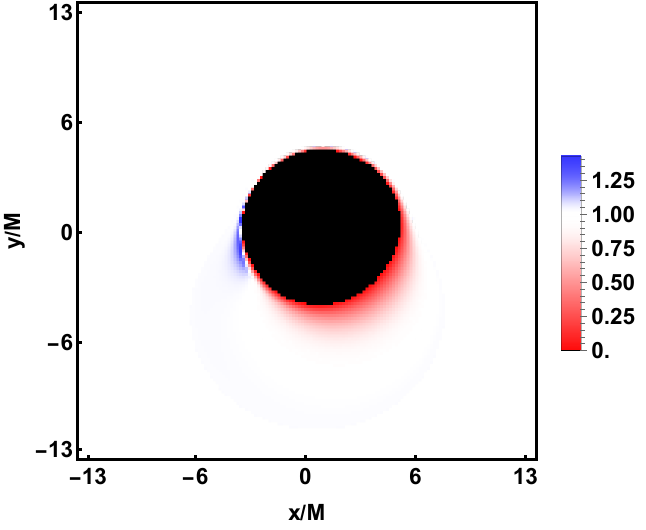}}
\subfigure[\tiny][~$\gamma=0.02,~\mu=0.02$]{\label{a1}\includegraphics[width=5.4cm,height=5.2cm]{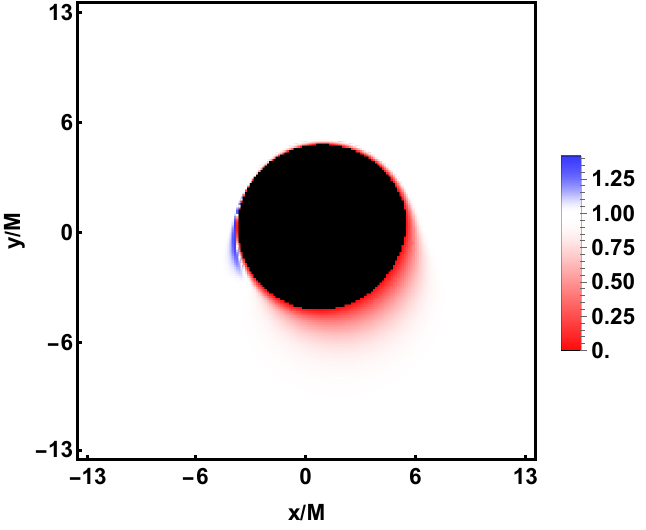}}
\subfigure[\tiny][~$\gamma=0.02,~\mu=0.05$]{\label{b1}\includegraphics[width=5.4cm,height=5.2cm]{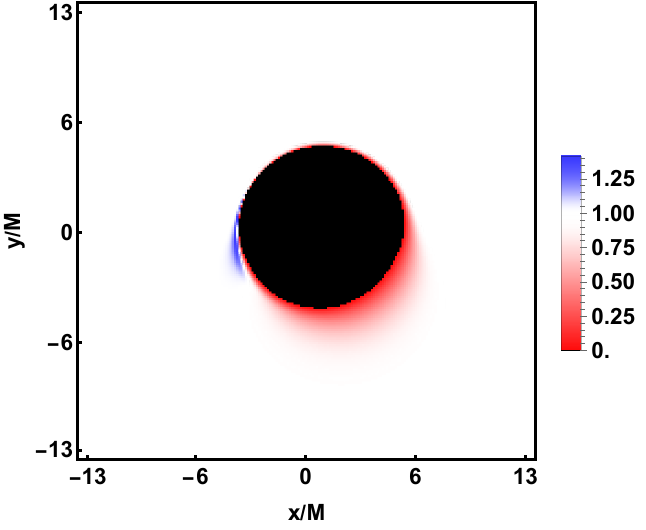}}
\subfigure[\tiny][~$\gamma=0.02,~\mu=0.1$]{\label{c1}\includegraphics[width=5.4cm,height=5.2cm]{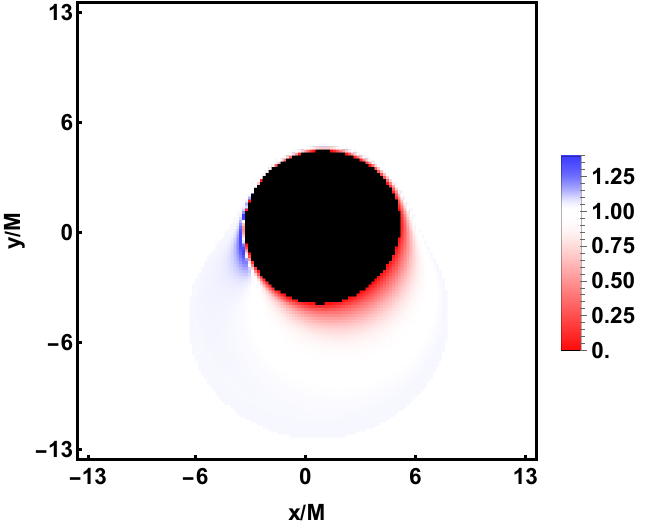}}
\subfigure[\tiny][~$\gamma=0.03,~\mu=0.02$]{\label{a1}\includegraphics[width=5.4cm,height=5.2cm]{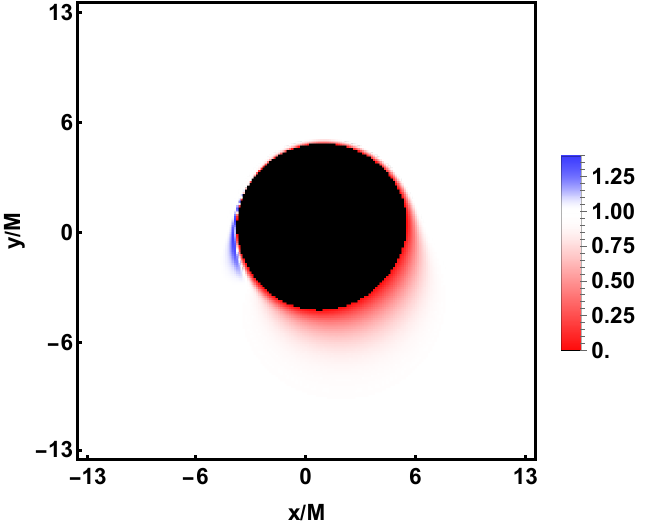}}
\subfigure[\tiny][~$\gamma=0.03,~\mu=0.05$]{\label{b1}\includegraphics[width=5.4cm,height=5.2cm]{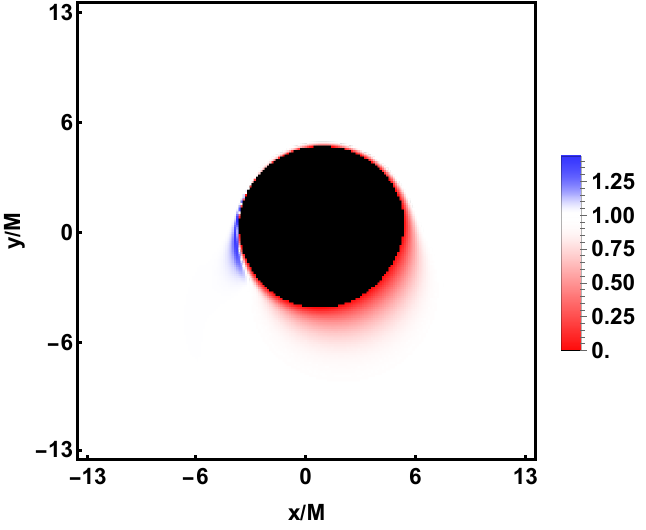}}
\subfigure[\tiny][~$\gamma=0.03,~\mu=0.1$]{\label{c1}\includegraphics[width=5.4cm,height=5.2cm]{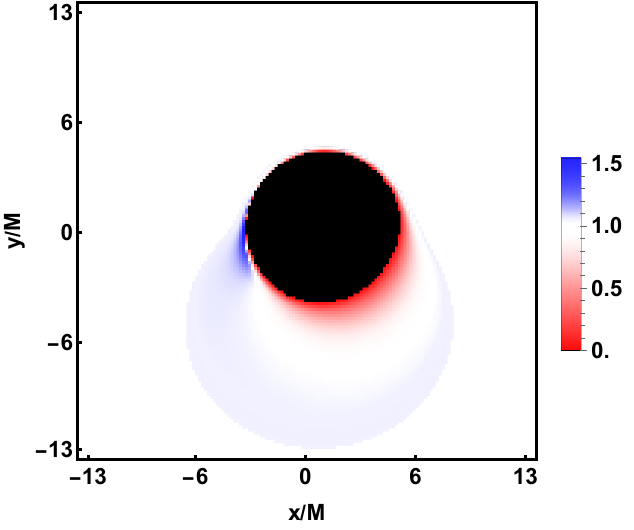}}
\caption{Redshift maps of the lensed images for a rotating BH in Rastall gravity under a prograde accretion flow shown for different values of $\mu$ and $\gamma$ with fixed $a=0.5$ and $\theta_{ obs}=70^\circ$. The red and blue regions correspond to redshift and blueshift photon emission, respectively, while the black shaded area denotes the BH event horizon.
}\label{prd7}
\end{figure}
Figure \textbf{\ref{prd7}} presents the redshift distribution of the lensed images for a rotating BH in Rastall gravity under a prograde accretion flow. In all panels, the lensed structures surrounding the shadow exhibit redshifted and blueshifted emission, with the latter becoming increasingly prominent as the model parameters vary. The central dark region remains enclosed by a bright lensed feature, while noticeable changes occur in the distribution of the redshift pattern across the image. For larger values of the parameters, particularly in the lower rows, the blueshifted emission extends over a wider area of the observer's screen and becomes comparable to, or even more pronounced than, the redshifted component. At the same time, the morphology of the lensed image undergoes visible modifications, indicating that the Rastall and structure parameters have a significant impact on the observed redshift configuration of the accretion flow.

\begin{figure}
\centering
\subfigure[\tiny][~$\gamma=0.01,~\mu=0.02$]{\label{a1}\includegraphics[width=5.4cm,height=5.2cm]{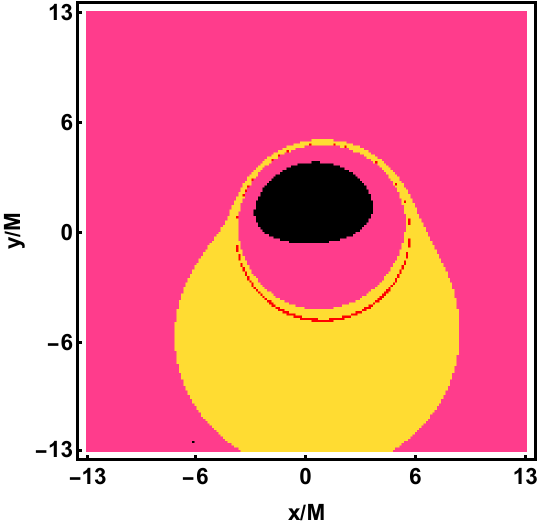}}
\subfigure[\tiny][~$\gamma=0.01,~\mu=0.05$]{\label{b1}\includegraphics[width=5.4cm,height=5.2cm]{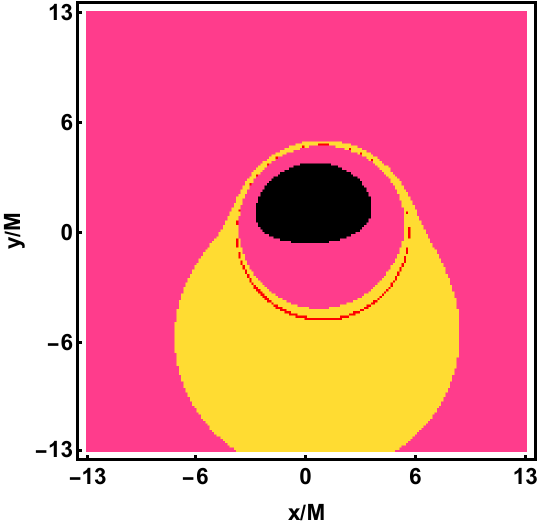}}
\subfigure[\tiny][~$\gamma=0.01,~\mu=0.1$]{\label{c1}\includegraphics[width=5.4cm,height=5.2cm]{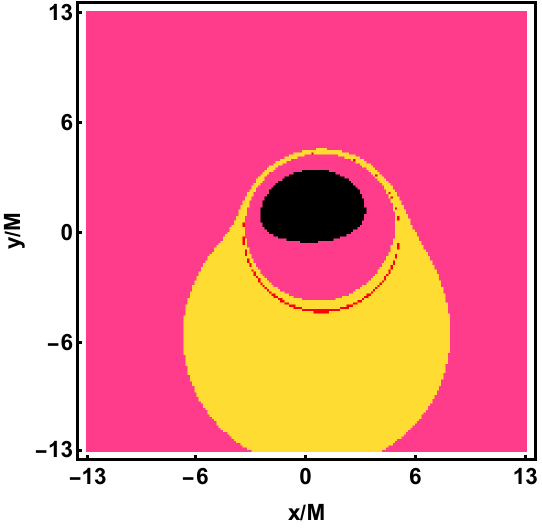}}
\subfigure[\tiny][~$\gamma=0.02,~\mu=0.02$]{\label{a1}\includegraphics[width=5.4cm,height=5.2cm]{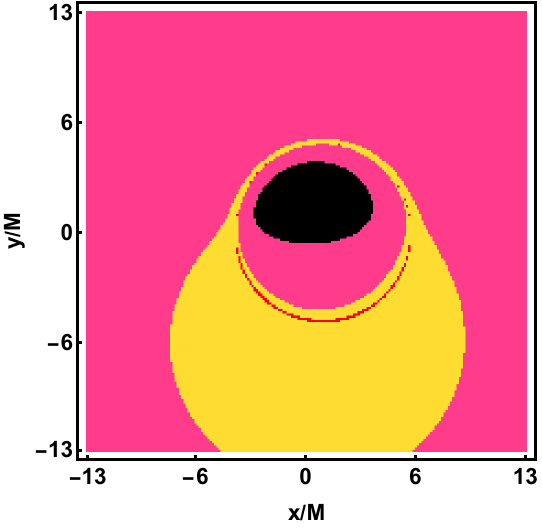}}
\subfigure[\tiny][~$\gamma=0.02,~\mu=0.05$]{\label{b1}\includegraphics[width=5.4cm,height=5.2cm]{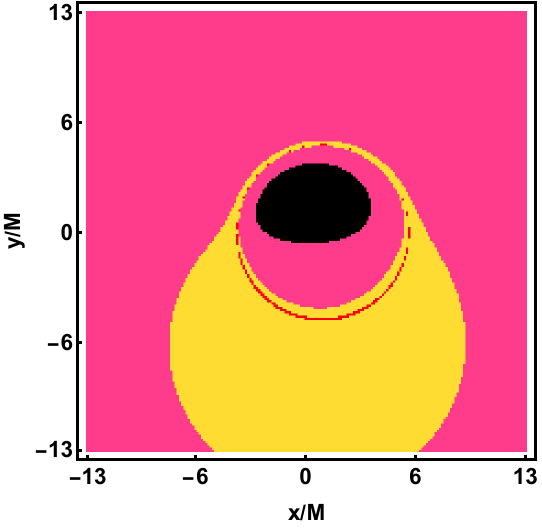}}
\subfigure[\tiny][~$\gamma=0.02,~\mu=0.1$]{\label{c1}\includegraphics[width=5.4cm,height=5.2cm]{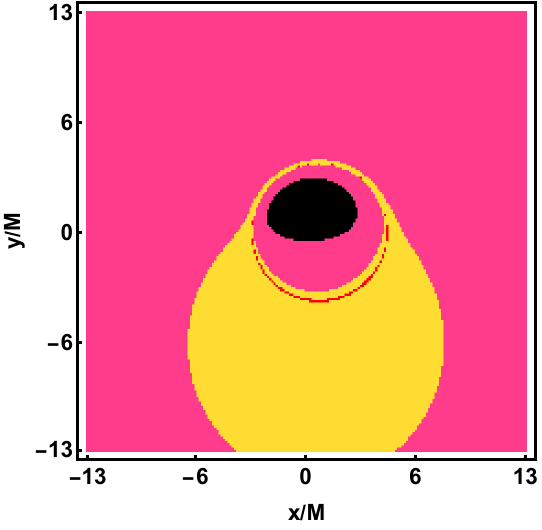}}
\subfigure[\tiny][~$\gamma=0.03,~\mu=0.02$]{\label{a1}\includegraphics[width=5.4cm,height=5.2cm]{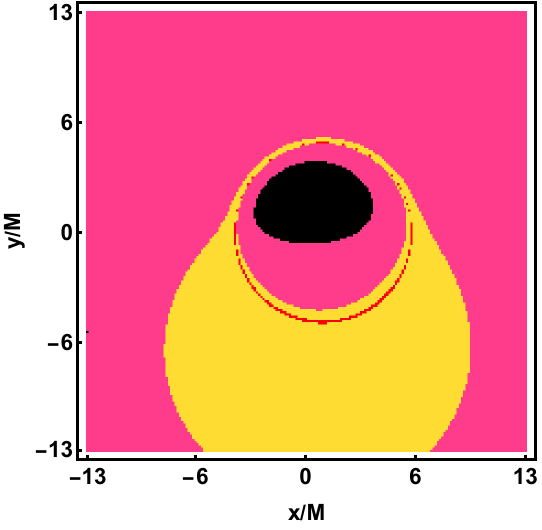}}
\subfigure[\tiny][~$\gamma=0.03,~\mu=0.05$]{\label{b1}\includegraphics[width=5.4cm,height=5.2cm]{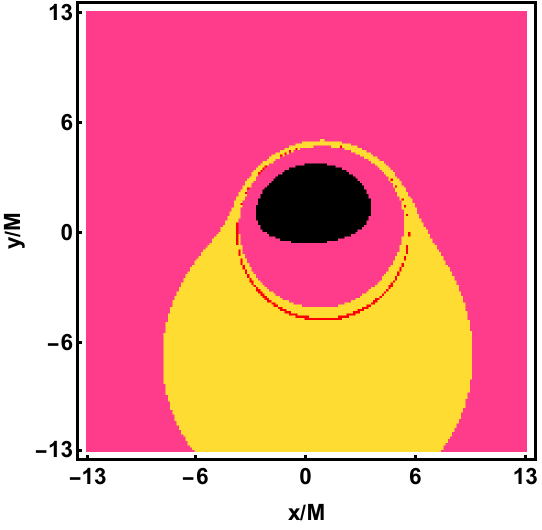}}
\subfigure[\tiny][~$\gamma=0.03,~\mu=0.1$]{\label{c1}\includegraphics[width=5.4cm,height=5.2cm]{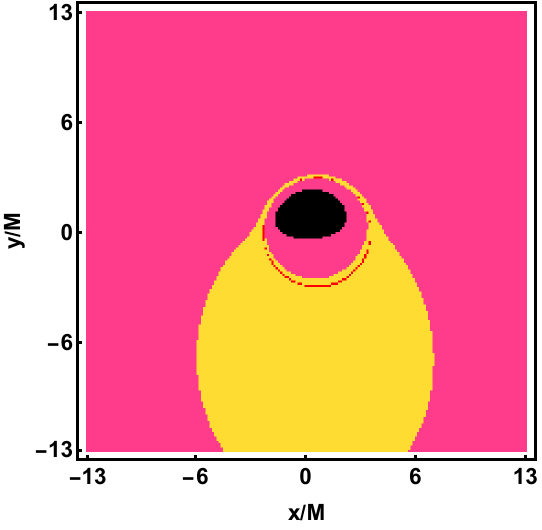}}
\caption{Lensing bands of a rotating BH in Rastall gravity under a prograde accretion flow for different values of $\mu$ and $\gamma$ with fixed $a=0.5$ and $\theta_{ obs}=70^\circ$. The pink, yellow, and red regions correspond to the direct, lensed, and photon ring images, respectively, while the black shaded area represents the BH event horizon.
}\label{prd8}
\end{figure}
To further distinguish direct and lensed images, the corresponding lensing bands are presented in the Fig. \textbf{\ref{prd8}} for the same parameter values considered previously. In all panels, the photon ring remains enclosed within the direct and lensed bands. The direct image occupies the largest portion of the screen, whereas the lensed band is mainly concentrated in the lower half of the image. As $\mu$ increases, the central dark region gradually decreases in size, while noticeable changes occur in the morphology of the lensed band. In particular, for the largest value of $\mu$, the lensed band becomes more confined and is reduced in width. A similar behavior is observed with increasing values of $\gamma$, which further modifies the overall lensing pattern. In contrast, the photon ring undergoes only minor variations and remains relatively stable throughout the parameter space.
\begin{figure}
\centering
\subfigure[\tiny][~$\gamma=0.02,~\mu=0.02$]{\label{a1}\includegraphics[width=5.4cm,height=5.2cm]{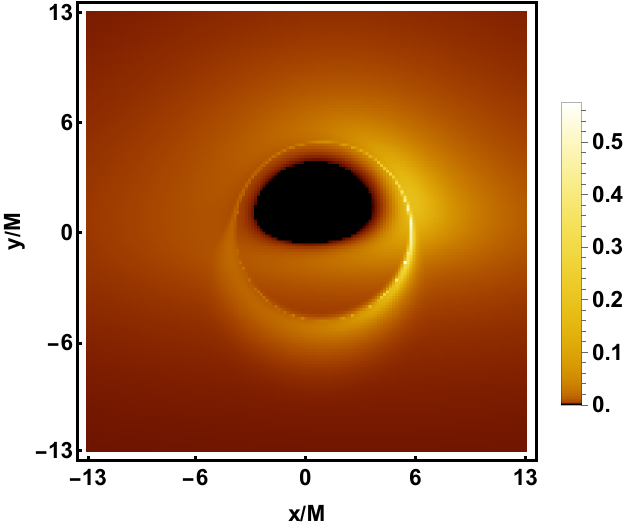}}
\subfigure[\tiny][~$\gamma=0.02,~\mu=0.05$]{\label{b1}\includegraphics[width=5.4cm,height=5.2cm]{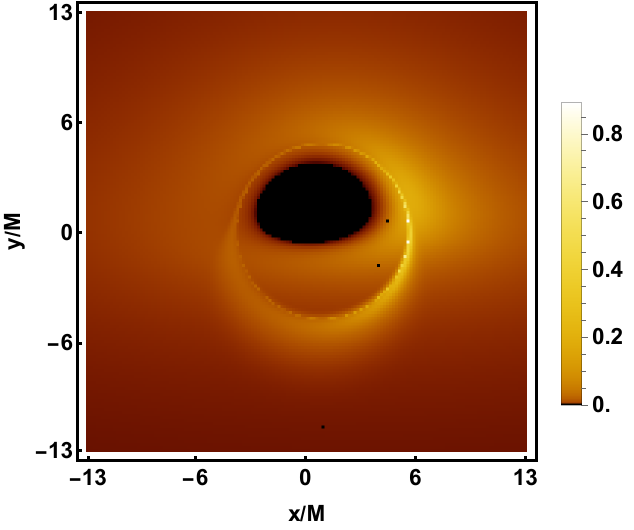}}
\subfigure[\tiny][~$\gamma=0.02,~\mu=0.1$]{\label{c1}\includegraphics[width=5.4cm,height=5.2cm]{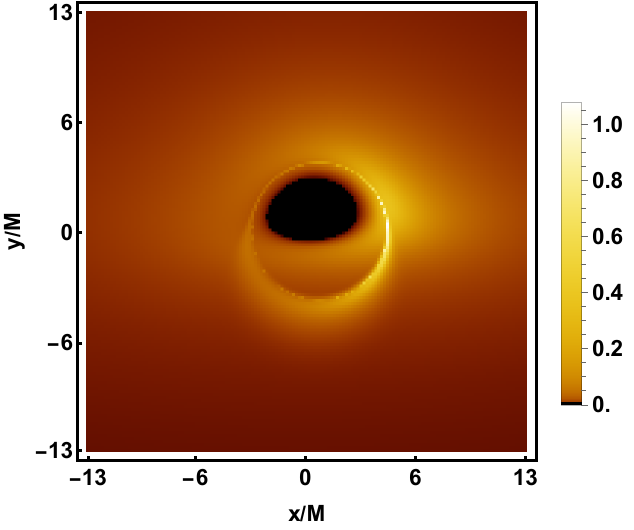}}
\caption{Optical images of rotating BH in Rastall gravity for different values of $\mu$ with fixed $\gamma=0.02$,~$a=0.5$ and $\theta_{obs}=70^\circ$ under a retrograde accretion flow. The BH's event horizon is depicted as a black region and a luminous circular ring corresponds to the position of the photon ring.}\label{prd9}
\end{figure}
The optical appearance of the rotating BH under a retrograde accretion flow are shown in Fig. \textbf{\ref{prd9}} for different values of $\mu$, while the spin parameter and structure parameter are fixed at $a=0.5$ and $\gamma=0.02$, respectively. As $\mu$ increases from left to right, the central shadow region gradually decreases in size, whereas the surrounding emission structure becomes more extended. In addition, the brightness distribution exhibits a noticeable asymmetry, reflecting the influence of the retrograde motion of the accreting matter. These results indicate that the Rastall parameter has a significant impact on the observed optical appearance of the BH and its surrounding accretion flow.

\begin{figure}
\centering
\subfigure[\tiny][~$\gamma=0.02,~\mu=0.02$]{\label{a1}\includegraphics[width=5.4cm,height=5.2cm]{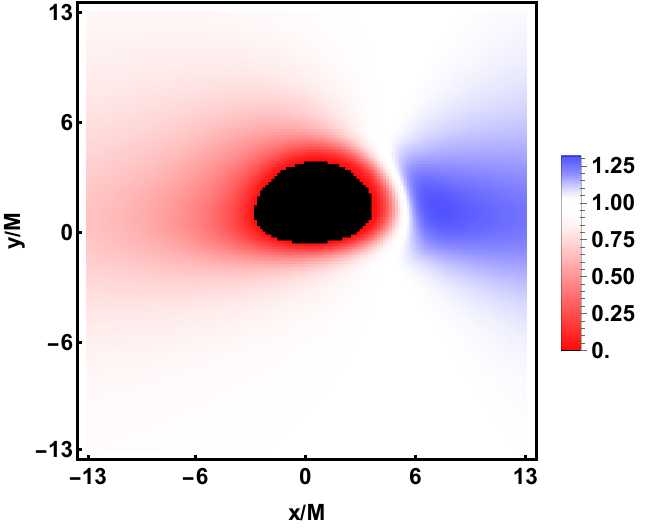}}
\subfigure[\tiny][~$\gamma=0.02,~\mu=0.05$]{\label{b1}\includegraphics[width=5.4cm,height=5.2cm]{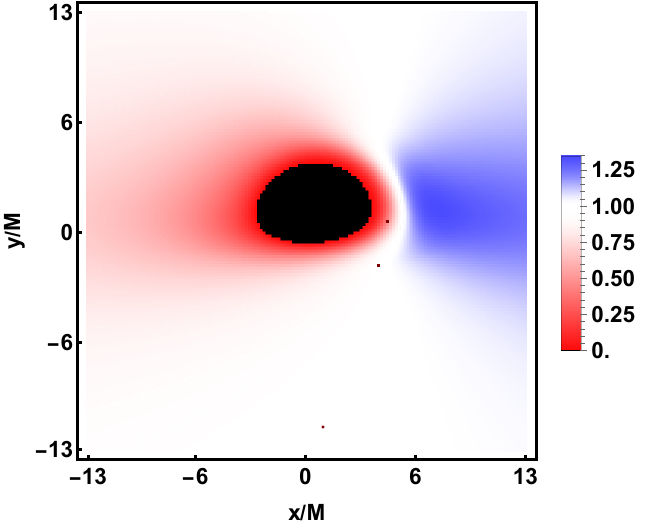}}
\subfigure[\tiny][~$\gamma=0.02,~\mu=0.1$]{\label{c1}\includegraphics[width=5.4cm,height=5.2cm]{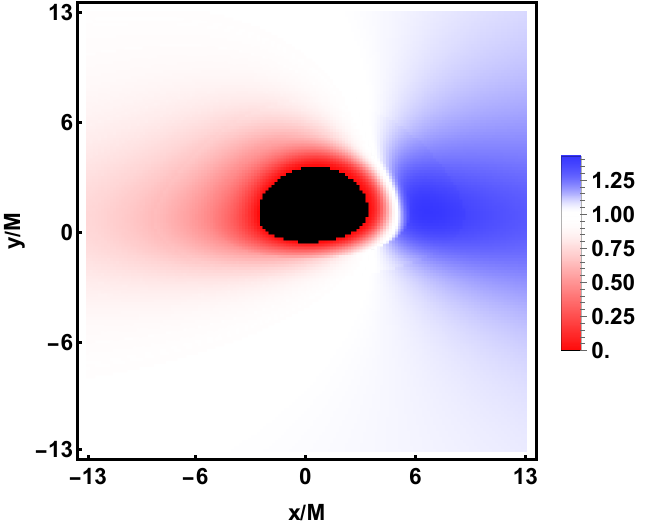}}
\caption{Redshift maps of the direct image for a rotating BH in Rastall gravity with a retrograde accretion flow for different values of $\mu$ and fixed $\gamma=0.02$, $a=0.5$, and $\theta_{ obs}=70^\circ$. The red and blue regions correspond to redshift and blueshift photon emission, respectively, while the black shaded area denotes the BH event horizon.
}\label{prd10}
\end{figure}
Figure \textbf{\ref{10}} shows the redshift distribution of direct images for a rotating BH in Rastall gravity under retrograde accretion flow. Here, $a=0.5$ and $\gamma=0.02$ are fixed, while $\mu$ increases from left to right. As $\mu$ increases, the blueshifted region expands and becomes dominant, particularly for $\mu=0.1$, where it occupies a significant portion of the image plane.

\begin{figure}
\centering
\subfigure[\tiny][~$\gamma=0.02,~\mu=0.02$]{\label{a1}\includegraphics[width=5.4cm,height=5.2cm]{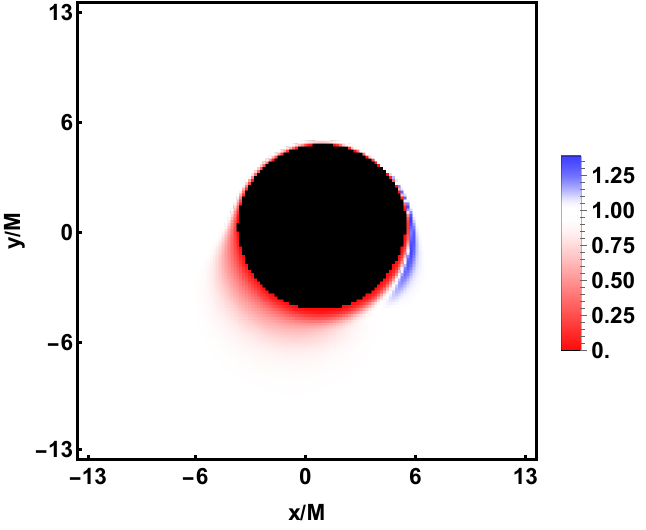}}
\subfigure[\tiny][~$\gamma=0.02,~\mu=0.05$]{\label{b1}\includegraphics[width=5.4cm,height=5.2cm]{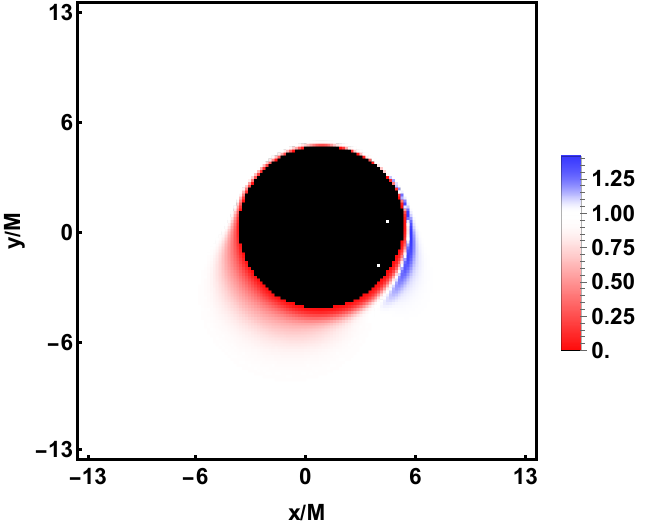}}
\subfigure[\tiny][~$\gamma=0.02,~\mu=0.1$]{\label{c1}\includegraphics[width=5.4cm,height=5.2cm]{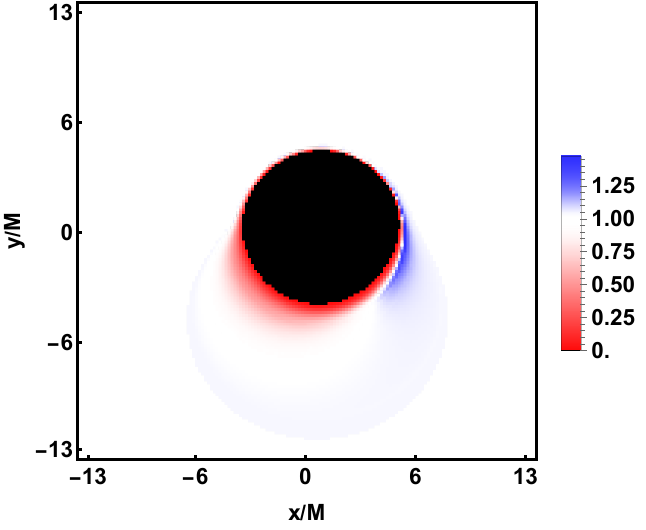}}
\caption{Redshift maps of the lensed images for a rotating BH in Rastall gravity with a retrograde accretion flow for different values of $\mu$ with fixed $\gamma=0.02$, $a=0.5$ and $\theta_{\rm obs}=70^\circ$. The red and blue regions correspond to redshift and blueshift photon emission, respectively, while the black shaded area denotes the BH event horizon.}\label{prd11}
\end{figure}
The redshift pattern of the lensed images under a retrograde accretion flow exhibits noticeable changes with increasing $\mu$ in Fig \textbf{\ref{11}}. The blueshifted emission gradually becomes more prominent and occupies a larger area of the image, while slight modifications appear in the surrounding lensed structures and the central dark region. These results highlight the influence of the Rastall parameter on the observed optical properties of the accretion flow.

\begin{figure}
\centering
\subfigure[\tiny][~$\gamma=0.02,~\mu=0.02$]{\label{a1}\includegraphics[width=5.4cm,height=5.2cm]{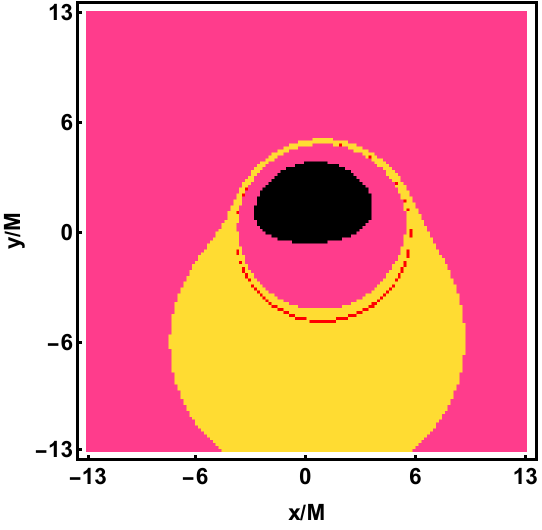}}
\subfigure[\tiny][~$\gamma=0.02,~\mu=0.05$]{\label{b1}\includegraphics[width=5.4cm,height=5.2cm]{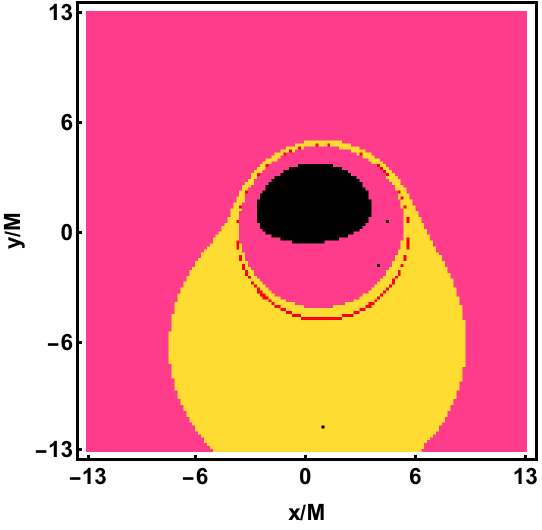}}
\subfigure[\tiny][~$\gamma=0.02,~\mu=0.1$]{\label{c1}\includegraphics[width=5.4cm,height=5.2cm]{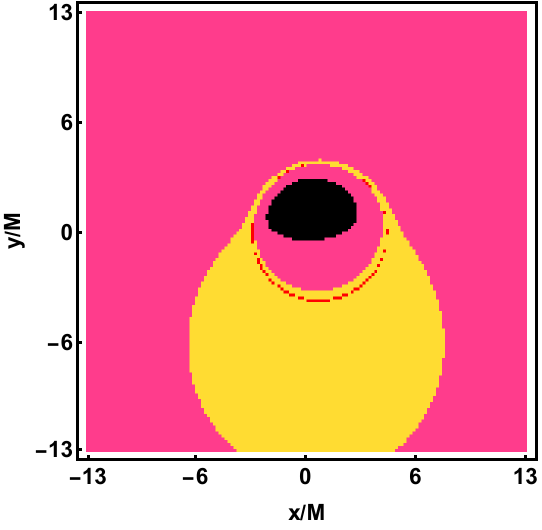}}
\caption{Lensing bands of a rotating BH in Rastall gravity under a retrograde accretion flow for different values of $\mu$ with fixed $\gamma=0.02$, $a=0.5$ and $\theta_{ obs}=70^\circ$. The pink, yellow, and red regions correspond to the direct, lensed, and photon ring images, respectively, while the black shaded area represents the BH event horizon.
}\label{prd12}
\end{figure}
The lensing bands for the retrograde accretion flow are shown in the Fig. \textbf{\ref{12}} with fixed $\gamma=0.02$ and $a=0.5$. As $\mu$ increases, the central dark region gradually shrinks, while the lensed band becomes narrower and more confined. In contrast, the photon ring remains nearly unchanged, indicating that the Rastall parameter primarily affects the lensing structure surrounding the BH.

\section{Constraints from EHT Observations}

In this section, we investigate the constraints on the model parameters of the rotating BH in Rastall gravity using the latest observational results from the EHT. The shadow measurements of M87$^\ast$ and Sgr A$^\ast$ provide a valuable framework for testing the viability of the model and examining the influence of $\mu$ and $\gamma$ on the BH shadow. To connect with the observations, we evaluate the angular diameter of the shadow and compare theoretical predictions with the EHT measurements of M87$^\ast$ and Sgr A$^\ast$. To assess the model's observational viability, the predicted shadow diameter is compared with the EHT measurements. A BH solution is regarded as compatible with the observational data when its angular diameter falls within the corresponding $1\sigma$ or $2\sigma$ confidence regions \cite{80}. The angular diameter $D$ of the shadow is given by
\begin{eqnarray}\label{27}
 D=\frac{2\overline{R}_d \hat{M}}{D_o},   
\end{eqnarray}
where $\overline{R}_d$ denotes the shadow radius evaluated at the BH position, $\hat{M}$ is the BH mass lies at distance $D_o$ from the observer. Consequently, the angular diameter can be expressed as \cite{80}
\begin{equation}\label{28}
D=2\times9.87098\overline{R}_{d}\big(\frac{\hat{M}}{M_{\odot}}\big)\big(\frac{1\text{kpc}}{D_{o}}\big)\mu as.
\end{equation}
The EHT observations of M87$^{\ast}$ and Sgr A$^{\ast}$ provide important benchmarks for testing BH models. For M87$^{\ast}$, the source distance and BH mass are $D_{o}=16.8\,\mathrm{Mpc}$ and $\hat{M}=(6.5\pm0.7)\times10^{6}M_{\odot}$, respectively, with an observed shadow diameter of $D_{\mathrm{M87}^{\ast}}=(37.8\pm2.7)\,\mu\mathrm{as}$ \cite{81}. For $\mathrm{Sgr\,A}^{*}$, the corresponding values are $D_{O}=8\,\mathrm{kpc}$, $\hat{M}=(4.0^{+1.1}_{-0.6})\times10^{6}M_{\odot}$, and $D_{\mathrm{Sgr\,A}^{\ast}}=(48.7\pm7)\,\mu\mathrm{as}$ \cite{82}.

\begin{figure}
\centering
\subfigure[\tiny][~$\mu=0.02$]{\label{a1}\includegraphics[width=7cm,height=7cm]{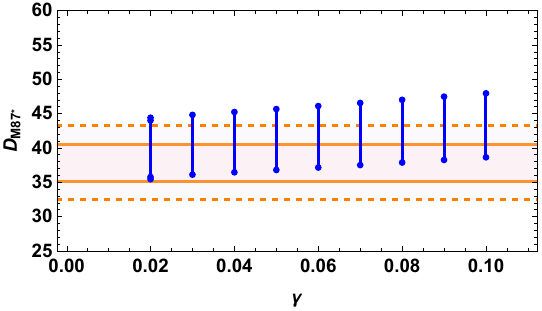}}
\subfigure[\tiny][~$\mu=0.05$]{\label{c1}\includegraphics[width=7cm,height=7cm]{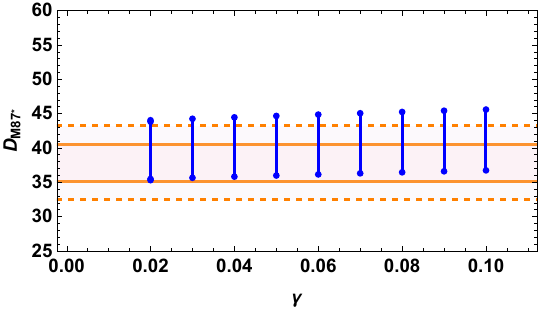}}
\subfigure[\tiny][~$\mu=0.02$]{\label{a2}\includegraphics[width=7cm,height=7cm]{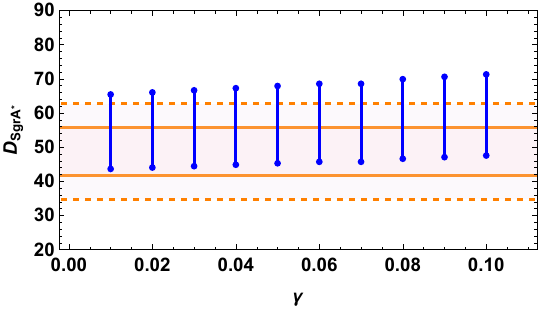}}
\subfigure[\tiny][~$\mu=0.05$]{\label{c2}\includegraphics[width=7cm,height=7cm]{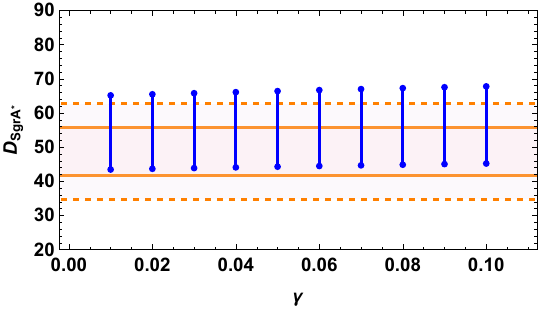}}
\caption{Comparison of the shadow angular diameter with the EHT observations for M87$^{\ast}$ (top row) and  Sgr A$^{\ast}$ (bottom row). The orange solid and dashed lines represent the observational $1\sigma$ and $2\sigma$ confidence levels, respectively. The blue segments indicate the parameter intervals consistent with the EHT measurements, while the corresponding bounds are marked by thick blue tick marks.}\label{prd13}
\end{figure}
Figure \textbf{\ref{prd13}} presents a comparison between the predicted shadow angular diameter $D$ and the observations of M87$^{\ast}$ (top row) and Sgr A$^{\ast}$ (bottom row). The structure parameter $\gamma$ is varied in all panels, while the spin parameter is fixed at $a=0.5$. The Rastall parameter is set to $\mu=0.02$ in the left panels and $\mu=0.05$ in the right panels. It can be seen that both the lower and upper bounds of the predicted $D$ increase gradually with increasing $\gamma$, indicating that the structure parameter enhances the apparent size of the BH shadow. For $\mu=0.02$, the variation of $D$ with $\gamma$ is more pronounced, leading to a wider interval of allowed $D$. When the Rastall parameter increases to $\mu=0.05$, the same increasing trend is preserved; however, the growth rate becomes slightly weaker, resulting in a comparatively narrower range of $D$. In both cases, the predicted values remain within the observational confidence levels provided by the EHT, demonstrating the consistency of the model with the shadow measurements of  M87$^{\ast}$.

A similar behavior is observed in the lower row for Sgr A$^{\ast}$. The angular diameter increases monotonically with $\gamma$ for both $\mu=0.02$ and $\mu=0.05$, showing that the influence of the structure parameter on the shadow size is independent of the astrophysical source. Nevertheless, the corresponding values of $D$ are systematically larger than those obtained for  M87$^{\ast}$, reflecting the different observational characteristics of Sgr A$^{\ast}$. The increase in $\mu$ again leads to a mild reduction in the spread of the predicted $D$, while preserving the overall trend with $\gamma$. Moreover, all predicted intervals remain compatible with the EHT observational bounds, indicating that the parameter space considered provides viable descriptions of both M87$^{\ast}$ and Sgr A$^{\ast}$. These results suggest that $\gamma$ plays a significant role in determining the shadow size, whereas $\mu$ mainly moderates the rate at which the $D$ varies.
\section{Final Remarks}
Thanks to the EHT for releasing the first image of the supermassive BH at the center of the Milky Way Galaxy, which opened a new observational window for investigating gravitational phenomena in the strong-field regime. These observations have stimulated considerable interest in examining how alternative theories of gravity may influence the appearance of BH shadows and the surrounding emission structures. In this context, we explored the optical properties of a rotating BH in Rastall gravity, focusing on the role of $\mu$ and $\gamma$ on the resulting BH shadow images.
To explore the observational consequences of the spacetime geometry, we analyzed the shadow contours and related observables under both celestial-sphere illumination and thin accretion disk environments. The effects of the model parameters were further examined through intensity maps, lensing bands, and redshift distributions for both prograde and retrograde accretion flows. In addition, the predicted shadow $D$ was compared with the EHT measurements of M87$^{\ast}$ and Sgr A$^{\ast}$, allowing the compatibility of the model with current observational data to be assessed.

Our results show that $\mu$ and  $\gamma$ leave noticeable imprints on the optical appearance of the rotating BH. The shadow analysis reveals that increasing either parameter enlarges the shadow size, while the deformation remains relatively weak, resulting in larger, nearly circular shadow profiles. This behavior is further supported by the shadow observables, where $R_d$ increases and $\delta_d$ decreases as $\gamma$ increases; see Fig. \textbf{\ref{prd3}}. From Fig. \textbf{\ref{prd4}}, we observe that the celestial-sphere images exhibit a similar trend, as larger values of $\mu$ and $\gamma$ lead to an expansion of the Einstein ring and the surrounding lensing structures, accompanied by a reduction in the size of the central dark region. Under thin accretion-disk illumination, the prograde and retrograde configurations exhibit a clear dependence on both $\mu$ and $\gamma$. The intensity maps show that increasing these parameters reduces the size of the central dark region while enhancing the surrounding emission structures, see Fig. \textbf{\ref{prd5}, \ref{prd9}}. The redshift distributions presented in the Figs. \textbf{\ref{prd6}} and \textbf{\ref{prd7}} for the prograde flow, and in \textbf{\ref{prd10}} and \textbf{\ref{prd11}} for the retrograde flow, show that the blueshift component becomes increasingly dominant with increasing $\mu$, occupying a progressively larger region of the observer's screen. At the same time, the extent of the redshift regions is gradually reduced, highlighting the significant influence of the $\mu$ on the observed frequency shifts of the emitted radiation. The lensing-band images shown in Figs. \textbf{\ref{prd8}} and \textbf{\ref{prd12}} reveal that increasing $\mu$ leads to a gradual reduction in the size of the central dark region, while the lensed structures undergo noticeable changes in shape and width. Finally, the comparison with the EHT observations of M87$^{\ast}$ and Sgr A$^{\ast}$ demonstrates that the predicted shadow $D$ remains compatible with the observational confidence intervals for a broad range of parameter values. 

Taken together, these results demonstrate that both $\mu$ and $\gamma$ play an important role in shaping the observable properties of rotating BHs. Their influence is reflected in the shadow geometry, emission profiles, redshift distributions, and lensing structures, leading to distinct observational signatures that may be tested with current and future high-resolution observations. The agreement of the predicted shadow $D$ with the EHT measurements of M87$^{\ast}$ and Sgr A$^{\ast}$ further supports the viability of the model within the considered parameter space. Therefore, rotating BH in Rastall gravity provides a promising framework for investigating deviations from GR and for exploring the nature of gravity in the strong-field regime. Of course, it is also interesting to check the influence of Rastall gravity in thick disk accretion images, polarized images, or hotspot images in future studies.

\end{document}